\documentclass{article}

\usepackage{arxiv}

\usepackage[utf8]{inputenc} 
\usepackage[T1]{fontenc}    
\usepackage{hyperref}       
\usepackage{url}            
\usepackage{booktabs}       
\usepackage{amsfonts}       
\usepackage{nicefrac}       
\usepackage{microtype}      

\usepackage{graphicx}

\usepackage{doi}
\usepackage{siunitx}
\usepackage{multirow}
\newcommand{\TabLLineLF}[1]{\includegraphics[width=.135\textwidth]{Am2Thz#1}&\includegraphics[width=.135\textwidth]{Pore2Thz#1}&\includegraphics[width=.135\textwidth]{NP2Thz#1}&\includegraphics[width=.135\textwidth]{Cone2Thz#1}&\includegraphics[width=.135\textwidth]{CT2Thz#1}&\includegraphics[width=.135\textwidth]{NW2Thz#1}\\}

\newcommand{\TabTLineLF}[1]{\includegraphics[width=.135\textwidth]{AmT2Thz#1}&\includegraphics[width=.135\textwidth]{PoreT2Thz#1}&\includegraphics[width=.135\textwidth]{NPT2Thz#1}&\includegraphics[width=.135\textwidth]{ConeT2Thz#1}&\includegraphics[width=.135\textwidth]{CTT2Thz#1}&\includegraphics[width=.135\textwidth]{NWT2Thz#1}\\}

\newcommand{\TabLLineHF}[1]{\includegraphics[width=.135\textwidth]{Am10Thz#1}&\includegraphics[width=.135\textwidth]{Pore10Thz#1}&\includegraphics[width=.135\textwidth]{NP10Thz#1}&\includegraphics[width=.135\textwidth]{Cone10Thz#1}&\includegraphics[width=.135\textwidth]{CT10Thz#1}&\includegraphics[width=.135\textwidth]{NW10Thz#1}\\}

\newcommand{\TabTLineHF}[1]{\includegraphics[width=.135\textwidth]{AmT4Thz#1}&\includegraphics[width=.135\textwidth]{PoreT5Thz#1}&\includegraphics[width=.135\textwidth]{NPT5Thz#1}&\includegraphics[width=.135\textwidth]{ConeT5Thz#1}&\includegraphics[width=.135\textwidth]{CTT5Thz#1}&\includegraphics[width=.135\textwidth]{NWT5Thz#1}\\}

\title{Ballistic Heat Transport in Nanocomposite: the Role of the Shape and Interconnection of Nanoinclusions}

\author{ \href{0000-0001-7257-8125}{\includegraphics[scale=0.06]{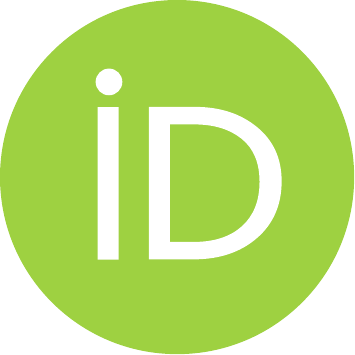}\hspace{1mm}Paul Desmarchelier} \\
	Univ Lyon, INSA Lyon, CNRS,CETHIL, UMR5008, LaMCoS, UMR5259\\
  	69621 Villeurbanne, France\\
	\texttt{paul.desmarchelier@insa-lyon.fr} \\
	\And
	Alice Carré \\
	Univ Lyon, INSA Lyon, CNRS, CETHIL, UMR5008, LaMCoS, UMR5259\\
	69621 Villeurbanne, France\\

	 \AND
 	\href{0000-0002-8521-7107}{\includegraphics[scale=0.06]{orcid.pdf}\hspace{1mm}Konstaninos Termentzidis} \\
	Univ Lyon, CNRS, INSA Lyon,CETHIL, UMR5008\\
	69621 Villeurbanne, France\\
	\texttt{konstantinos.termentzidis@insa-lyon.fr} \\
	\And
	Anne Tanguy\\
	Univ Lyon, INSA Lyon, CNRS, LaMCoS, UMR5259\\
	69621 Villeurbanne, France\\
	ONERA, University Paris-Saclay\\
	Chemin de la Hunière, BP 80100, 92123 Palaiseau, France\\
	\texttt{anne.tanguy@insa-lyon.fr} \\

}

\renewcommand{\headeright}{Submitted to Nanomaterials}
\renewcommand{\shorttitle}{ NC: Role of Shape and Interconnection of NIs}

\hypersetup{
pdftitle={Ballistic Heat Transport in Nanocomposite: the Role of the Shape and Interconnection of Nanoinclusions},
pdfsubject={cond-mat, nanomaterials},
pdfauthor={Paul Desmarchelier , Alice Carré, Konstantinos Termentzidis and Anne Tanguy },
pdfkeywords={Nanoinclusion, ballistic transport},
}

\begin{document}
\maketitle

\begin{abstract}
The effect on the vibrational and thermal properties of gradually interconnected nanoinclusions embedded in an amorphous silicon matrix is studied using MD simulations.
The nanoinclusion arrangement ranges from an aligned sphere array to an interconnected mesh of nanowires. Wave-packet simulations scanning different polarizations and frequencies reveal that the interconnection of the nanoinclusions at constant volume fraction induces a strong increase of the mean free path of high frequency phonons, but does not affect the energy diffusivity. The mean free path and energy diffusivity are then used to estimate the thermal conductivity, showing an enhancement of the effective thermal the effective thermal conductivity due to the existence of crystalline structural interconnections. This enhancement is dominated by the ballistic transport of phonons. Equilibrium molecular dynamics simulations confirm the tendency although less markedly. This leads to the observation that coherent energy propagation with a moderate increase of the thermal conductivity is possible. 
\end{abstract}


\section{Introduction}

Many applications in electronics require materials with tailored mechanical, electronic or thermal properties. To this end, the appropriate element, alloy, phase, crystallinity or a combination of them can be chosen. Nanostructuration allows a further improvement of performances. A wide variety of nanocomposites exists, one of the simplest consists of nanoinclusion (NIs) of a different phase or material embedded in a host matrix. 
Crystalline NIs in a crystalline matrix are used for many applications, such as thermoelectric generation~\cite{Kim2006}. For the same application crystalline NIs in an amorphous matrix have also been proposed~\cite{Zhu2007}. This last possibility takes advantage of the low thermal conductivity of the amorphous matrix while retaining some electronic transport properties of the added crystal.

However, the NIs and matrix influence each others~\cite{Murray2004,Damart2015}, notably their vibrational and thermal properties. A better understanding of the interaction of nanoinclusions and matrix is crucial to further improve the performances of these nanocomposites.

To study the heat dissipation through an amorphous/crystalline nanocomposite, one should understand both the physics of the amorphous material and this of the crystalline nanoclusters. The modern understanding of thermal transport in glasses was laid by Allen and Feldman~\cite{Allen1993,Allen1999}. They have introduced an intermediary transport regime between the localization and the propagation of vibrational modes: the diffusive regime. They have established a distinction between propagative and non-propagative modes. 	
In the former the phonon gas model can be applied, but not in the latter, due to strong scattering. Yet, some non-propagative modes still contribute to the thermal conductivity through energy diffusion. However, the distinction between propagative and diffusive modes is still under discussion; some authors argue that each mode has to be distinguished individually~\cite{Allen1993,Seyf2016} and others use a frequency limit to discriminate between propagative and diffusive modes~\cite{Beltukov2013,Larkin2014}. The used frequency limit is often set by the Ioffe-Regel criterion. This criterion relies on the comparison between the mean free path (MFP) and the wavelength~\cite{DeAngelis2019}. The sometimes blurred boundary between propagative and diffusive modes has led other authors to claim that a clear distinction between the two is not meaningful~\cite{Saaskilahti2016}.

The introduction of NIs in a solid matrix modifies the behavior of the material. For instance, a particle array can act as a low pass filter, scattering the higher frequencies~\cite{Damart2015}.
Different parameters have different effects: the rigidity contrast impacts the scattering and eventually pins the energy~\cite{Luo2019}. A higher surface to volume ratio is known to decrease the effective thermal conductivity~\cite{Termentzidis2018}. Less instinctively, it has been shown in the same study that, the relative crystalline orientation between the particles also modifies the thermal conductivity of the material, banning or promoting the phonon percolation. The size distribution of the NIs has also been proposed to reduce the thermal conductivity of crystal-crystal nanocomposites~\cite{Zhang2015}. Finally, the presence of NIs can cause an anticipation of the transition from propagative to diffusive regime in amorphous/crystalline nanocomposites~\cite{Tlili2019}. 

Most of the theoretical studies of NIs impact on the vibrational and thermal properties assume that NIs are spherical~\cite{Damart2015,Termentzidis2018,Tlili2019,Luo2019}. However, NIs can have multiple shapes~\cite{Hofmeister2005}. 
The shape influences the properties, for instance, NIs with a high surface to volume ratio increase the electrical conductivity in polymers~\cite{Vasudevan2019}. This ratio similarly increases the heat transport in nanofluids~\cite{Jabbari2017}. Moreover, when the mass fraction of NIs is high enough, the NIs can form a percolating network~\cite{Miura2015}. For Si NIs in a SiO$_2$ matrix the percolation can be controlled and modifies the properties of the material~\cite{Nakamura2015,Nakamura2018}.

A percolating network of NIs is similar to a nanomesh embedded in an amorphous matrix. Embedded NW meshes are already used in polymers to increase their thermal conductivity ~\cite{Huang2017a}. More generally, Car \textit{et al.} showed that it is possible to obtain single crystalline nanowire meshes (NW-M)~\cite{Car2014}. These NW-M, in 2D or 3D, are also known to have a low thermal conductivity compared to bulk material~\cite{Ma2016,Verdier2018e}.
Finally, a crystalline/amorphous nanocomposite is comparable to nanocrystalline materials. For these materials, studies exist about the transmission of phonons, across a single interface~\cite{Kimmer2007} or across multiple grain boundaries~\cite{Yang2017b}. The grain-size and grain-size distribution also impact the transport~\cite{Hori2015}.
The purpose of the following paper is to gain a better understanding of the effect of the gradual interconnection of crystalline NIs on thermal conductivity and ballistic transport.

To this end, several structures are studied, using equilibrium molecular dynamics (EMD) to compute their thermal conductivity and wave packet propagation method to distinguish propagative and diffusive contribution.
After the description of the configurations used, the different analysis methods will be presented. First the qualitative impact of the inclusion at different frequencies will be considered, then the vibrational properties of the different configurations will be studied. These properties are used to estimate the thermal conductivity, via the kinetic theory of gases framework, and finally this thermal conductivity will be compared to the results obtained with the EMD methodology. Finally, the impact of ballistic transport and NIs interconnection on the effective thermal conductivity are discussed.

\section{Materials and Methods}\label{sec:methode}

\subsection{Studied Configuration} \label{subsec:ConfCrea}

The nanocomposites studied here are composed of crystalline Si NIs embedded in an amorphous Si matrix. The NIs shapes and interconnections are varied to study their impact on the effective thermal conductivity and on the ballistic transport. The NIs are gradually interconnected, from an array of spherical NIs to a 3D nanowire mesh. The host matrix is an amorphous Si cube of side 11.9~nm cut out of a larger sample obtained in a previous study~\cite{Tlili2019}. This length is adapted to get an integer number of crystalline primitive cells and thus a monocrystal in case of structural percolation. Periodic boundary conditions are used in all directions. 

The nanocomposites are formed in the following manner: the NI shape is first hollowed out of the matrix, and then filled by crystalline Si (c-Si). The added crystal has the $\langle 100 \rangle$ direction aligned with the $x$ axis. 
In order to avoid the superposition of atoms when the crystalline phase is added, the holes are larger than the NIs themselves by .1 ~\si{\angstrom}. The created NIs have the same volume, so that all configurations have the same crystalline fraction (30\% of crystalline phase overall). Four shapes of NIs are considered, a sphere (S) with a radius of 5~nm (see Table~\ref{fig:Conf} second column); a sphere with six conical extremities pointing in the Cartesian coordinate directions without reaching the edges of the simulation box (see Table~\ref{fig:Conf} third column) it will be referred as sphere with cones (SC); the third is similar to the former but has longer conical extremities that reaches the simulation cell boundaries (see Table~\ref{fig:Conf} fourth column) it will be referred as sphere with truncated cones (STC); and finally a 3D crossing of three nanowires of 2.5~nm in radius aligned with the Cartesian coordinates (see Table~\ref{fig:Conf} last column) it will be referred as nanowire mesh (NW-M). This is around this last NI shape that the box size was set. All NIs are centered in the simulation box. For the SC configuration, the central sphere has a radius of 4.6~nm and the added cones have an opening angle of \ang{40}.

The apex of the cones are 0.3 nm away from the simulation box edges and the basis of the cones is prolonged until it intersects with the central sphere. This results in a neck of 0.6~nm between 2 inclusions.
For the STC configuration, the cones have a radius of 1~nm at their junction with the box boundary and are prolonged with an opening angle of \ang{34} until the intersection with the central sphere of radius 3.7~nm. Thus, only the STC and the NW-M have a continuous crystalline path across their simulation box, this continuous crystalline path across the structure will be referred to as crystalline structural percolation. This structural percolation has a minimum diameter of 2 nm for the STC and of 5 nm for the NW-M. Additionally, we have studied a porous sample, with spherical pores of the same diameter as the S system (see Table~\ref{fig:Conf} first column), and a fully amorphous sample is also studied for the sake of comparison. In Table~\ref{fig:Conf}, the different NIs are represented in 3D in the first row and in the second row a cross-section at the middle of the corresponding nanocomposite is depicted. These representations are obtained thanks to OVITO~\cite{Stukowski2010}.
\begin{table}[h]
	\caption{\label{fig:Conf} Freestanding nanoparticles (first row) and a cross-section of NIs embedded in an a-Si matrix (in dark gray) cross-section (second row). In each case the NI represent 30\% of the volume of simulation cell.}
	\centering
	\begin{tabular}{ccccc}
		\toprule
		\textbf{Pore}&\textbf{Sphere}&\textbf{SC}&\textbf{STC}&\textbf{NW-M}\\
		\midrule
		\multirow{2}{*}[+18pt]{\includegraphics[width=2 cm]{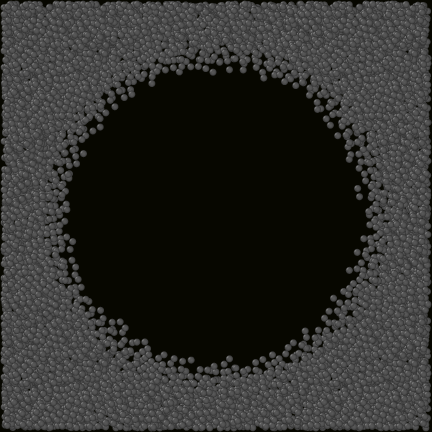}}&\includegraphics[width=2 cm]{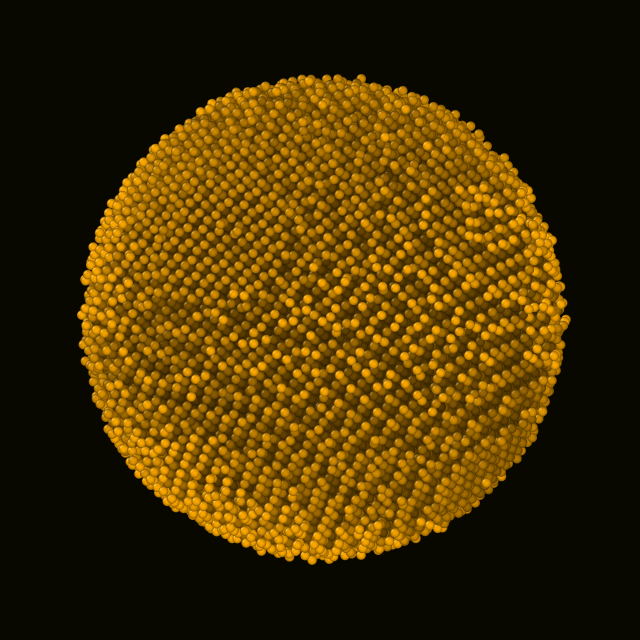}&\includegraphics[width=2 cm]{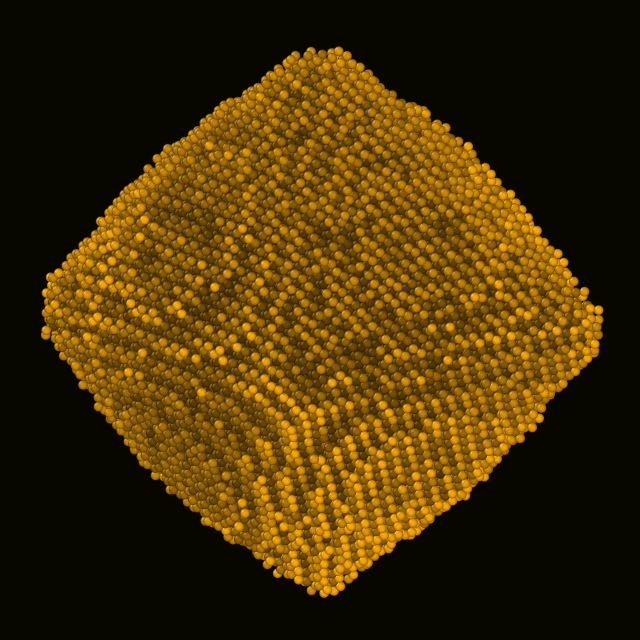}&\includegraphics[width=2 cm]{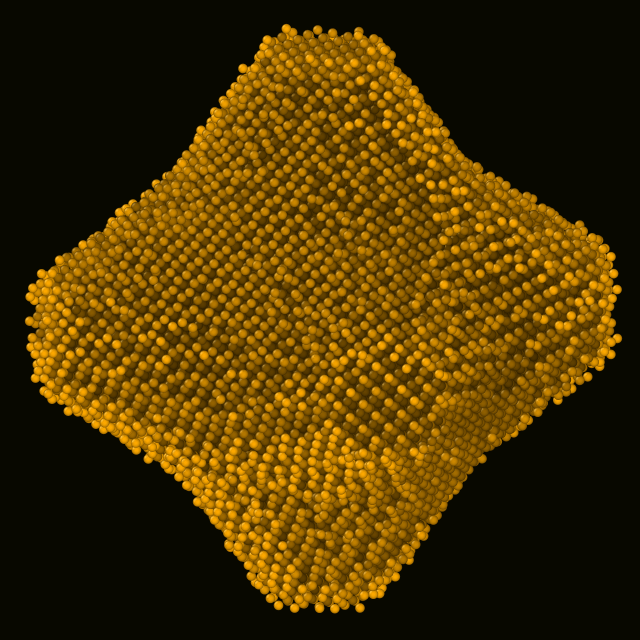}&\includegraphics[width=2 cm]{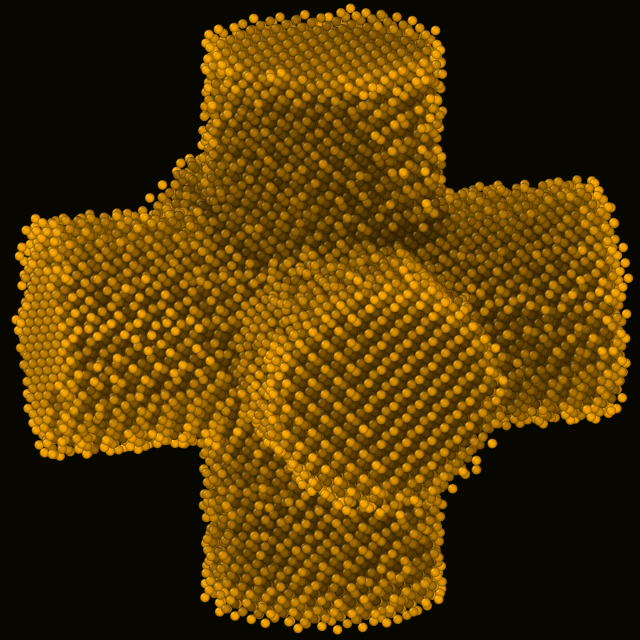}\\
		&\includegraphics[width=2 cm]{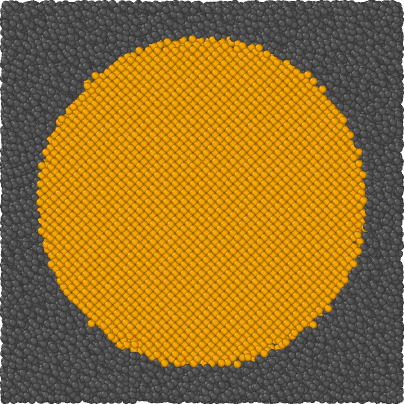}&\includegraphics[width=2 cm]{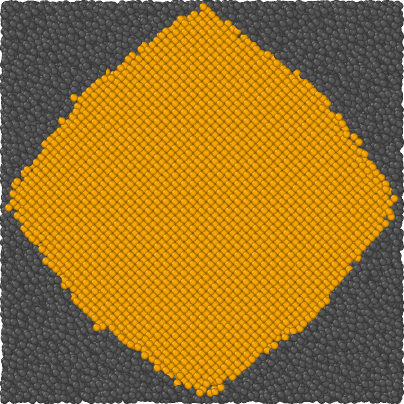}&\includegraphics[width=2 cm]{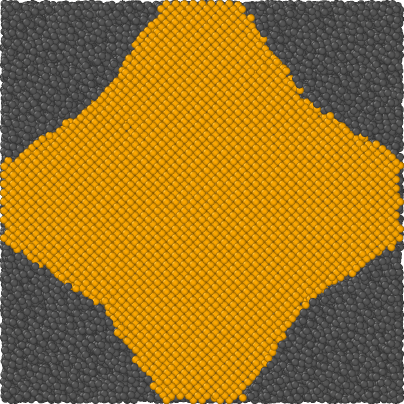}&\includegraphics[width=2 cm]{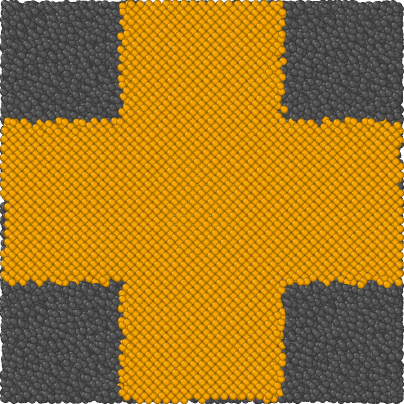}\\
		\bottomrule
	\end{tabular}
\end{table}

After the geometrical construction, the different simulations are annealed in the following manner. The atomic positions are relaxed using a conjugated gradient (CG) method, then the system is annealed at 100~K for 1~ps and finally a second conjugated gradient force minimization is performed\footnote{At this point, we want to mention that the relaxation induces small reconfigurations at the surface of the NIs. This interfacial reconfiguration causes a reduction of up to 3\% of the crystalline fraction and can change slightly the surface to volume ratio of the NIs. To take this into account, when computing the surface to volume ratio, only the particles recognized as diamond structure and first and second neighbors by a modified common neighbor analysis \cite{Maras2016} are considered.}. All modeling and MD simulations are carried out using the open-source software LAMMPS~\cite{Plimpton1997}.

We used a modified Stillinger-Weber (SW) potential~\cite{Vink2001} for its more realistic modeling of the interfaces between c-Si and a-Si in terms of interfacial energy and of atomic energies inside the two phases~\cite{France-Lanord2014}.

\subsection{Equilibrium Molecular Dynamics}
The Equilibrium Molecular Dynamics (EMD) method is used to estimate the thermal conductivity ($\kappa$) of the configurations previously described. This method relies on the fluctuation dissipation theorem linking the decay of the fluctuation of an internal variable to its response function. Here, the flux auto-correlation integral is linked to the thermal conductivity using the Green-Kubo formula~\cite{Kubo1947}:

\begin{equation}
\kappa_{\alpha\beta}=(Vk_BT^2)^{-1}\int_{0}^{\infty}\langle J_\alpha(0) J_\beta(t) \rangle dt
\end{equation} 

with $\alpha$ and $\beta$ the directions, $V$ the volume of the system, $k_b$ the Boltzmann constant and $J_\beta(t)$ the thermal flux in the direction $\beta$ at a time $t$ computed. The thermal flux is computed by LAMMPS using atomic energy for the convective part and the "group" atomic stresses for the virial contribution~\cite{Surblys2019}. A discretized version is~\cite{Schelling2002}:

\begin{equation}
\kappa_{\alpha\beta}=\Delta t(Vk_BT^2)^{-1}\sum_{m=1}^{M}(m-p)^{-1}\sum_{n=1}^{p} J_\alpha(m+n) J_\beta(m) 
\label{eq:GKdis}
\end{equation} 

with $\Delta t$ the time step between two successive flux computations, $M$ the total number of time steps, and $p$ the number of time step over which the auto-correlation function is averaged.

Before computing the flux auto-correlation function, the configurations (one inclusion in the amorphous cube as represented in the bottom row of Table~\ref{fig:Conf}) are first heated at 50~K, using a random initial velocity distribution. 
After that, the temperature is increased from 50 to 600~K at constant pressure in 0.05~ns. Then the system is annealed at 600~K with a Nos\'e Hoover thermostat for 0.25~ns to ensure better temporal stability. To insure the absence of recrystallization, it is checked that this annealing does not impact the radial distribution function at 300K. After this annealing, the temperature is decreased to 300~K at constant pressure in 0.05~ps and then equilibrated at 300~K for 2~ns. 
The flux auto-correlation function is finally measured during 10~ns in a constant energy simulation. For all the simulations a time step of \num{5e-7}~ns is used. For the computation of the auto-correlation, the flux is sampled every \num{1e-5}~ns and the flux auto-correlation decay is computed over 0.04~ns. These simulations are repeated 5 times, with a different initial velocity distribution for each repetition, to get better statistics. The final value is the mean $\kappa$ across the simulations and the uncertainty range is defined by the highest and lowest values of the individual runs. 
\subsection{Thermal Conductivity from the Kinetic Theory} 

The thermal conductivity of the different configurations can also be evaluated using their vibrational properties, for this the method initially developed by Tlili \textit{et al.}~\cite{Tlili2019} for spherical NIs is used. The contribution of the propagative and the diffusive modes are separated. The propagative contribution ($\kappa_{P}$) is estimated with the following integral~\cite{Ashcroft2002}:

\begin{equation}
\kappa_{P}=\sum_\eta\int^{\nu_{max}}_0 m_{\eta} C(\nu,T) v^2_{\eta}(\nu)\tau_{\eta}(\nu) g_{\eta}(\nu) d\nu
\label{eq:kp}
\end{equation}

with $C(\nu,T)$ the heat capacity at the temperature $T$ and frequency $\nu$, $v_{\eta}(\nu)$ the group velocity, $\tau_{\eta}(\nu)$ the phonon lifetime, $g_{\eta}(\nu)$ the density of state at the frequency $\nu$, and $m_{\eta}$ the degree of freedom associated to the polarization $\eta$ (longitudinal or transverse). $\nu_{max}$ is the frequency for which the group velocity is zero or ill-defined.

The contribution of the diffusive part ($\kappa_{D}$) can be estimated through~\cite{Larkin2013}:

\begin{equation}
\kappa_{D}= \int^{\nu_{maxD}}_{0} 3 C(\nu,T) g(\nu)D(\nu) d\nu
\label{eq:kd}
\end{equation}

with $D(\nu)$ the diffusivity at the frequency $\nu$. $\nu_{maxD}$ is the frequency at which the diffusivity is considered negligible, that is 15~THz for the configurations studied. The heat capacity, using the Debye Model~\cite{Ashcroft2002}, is given as follows:

\begin{equation}
C(\nu,T)=\frac{Nk_b}{V}(\frac{2\pi\nu \hbar}{k_bT})^2\frac{\exp{(\frac{2\pi\nu \hbar}{k_bT})}}{(\exp{(\frac{2\pi\nu \hbar}{k_bT})}-1)^2}
\label{eq:C}
\end{equation}

with $k_b$ the Boltzmann constant, $\hbar$ the Planck constant, $V$ the volume, and N the number of atoms. The method for the estimation of other components of Equations~\ref{eq:kp} and~\ref{eq:kd} are detailed in section \ref{sec:WP}.

The global thermal conductivity will be taken as the sum of the diffusive and propagative contributions. Here, as both propagative and diffusive behaviors appear at most frequencies, both contributions are considered over the whole spectrum.

\subsection{Wave Packet Propagation}
\label{sec:WP}
The wave packet (WP) method is used to study the different aspects of the phononic contribution to the thermal conductivity~\cite{Beltukov2016}. This method enables the estimation of the Mean Free Path (MFP) and diffusivity in a dual wave/particle description of phonons. These quantities are estimated thanks to the excitation of different vibrational modes and the measure of their decay rate according to space and time.

First, the media in which the WP propagates is obtained from the repetition of the cubes described in section~\ref{subsec:ConfCrea}. They are repeated 6 times in the $x$ direction. Indeed, a sufficiently long sample is needed to the study the spatial decay of the WP. Before the excitation, the atomic velocities are set to 0 and the position of the atoms are relaxed using a CG method to minimize the force so that any movement of the atoms will be caused by the WP. Then, an excitation is applied in a central slice of 0.2~nm between two repetitions of the initial configuration. This excitation is a Gaussian windowed sinusoidal force impulsion,

\begin{equation}
f=A \sin[ 2 \pi\nu(t-3\tau)]\exp{\left[-\frac{(t-3\tau)^2}{(2\tau^2)}\right]}.
\label{eq:foreeq}
\end{equation}

The amplitude $A$ is chosen to be sufficiently low to avoid anharmonic effects (here \num{3.773 e-4}\si{\eV\per\angstrom}). The Gaussian window width $\tau$ balances between spatial extension of the WP compared to the nanocomposite length, and the resolution in the frequency space, here \num{36e-4}~ns. The studied frequencies range from 1~THz, which is the limit of the resolution due to the $\tau$ used, to 15~THz by increments of 1~THz.

The force $f$ can be applied parallel to the principal dimension, creating longitudinal (L) waves or perpendicular to it creating transverse (T) waves. Alternatively, the force can be applied in a random direction, different for each atom, preventing the formation of a coherent wave. This random excitation with a uniform angle distribution is used to compute the energy diffusivity~\cite{Beltukov2013}.\sisetup{round-precision = 2} After the impulsion, the kinetic energy as a function of position over the $x$ axis is recorded every \num{1e-5}~ns from the creation of the impulsion until the wave fronts reach the periodic boundaries.\sisetup{round-precision = 1} The resolution along $x$ is of 0.72~nm. Additionally, the position and kinetic energy of every atom are recorded every \num{3e-4}~ns, in order to get a spatially resolved energy distribution.

\begin{figure}[h]
	\centering
	\includegraphics[width=.7\textwidth]{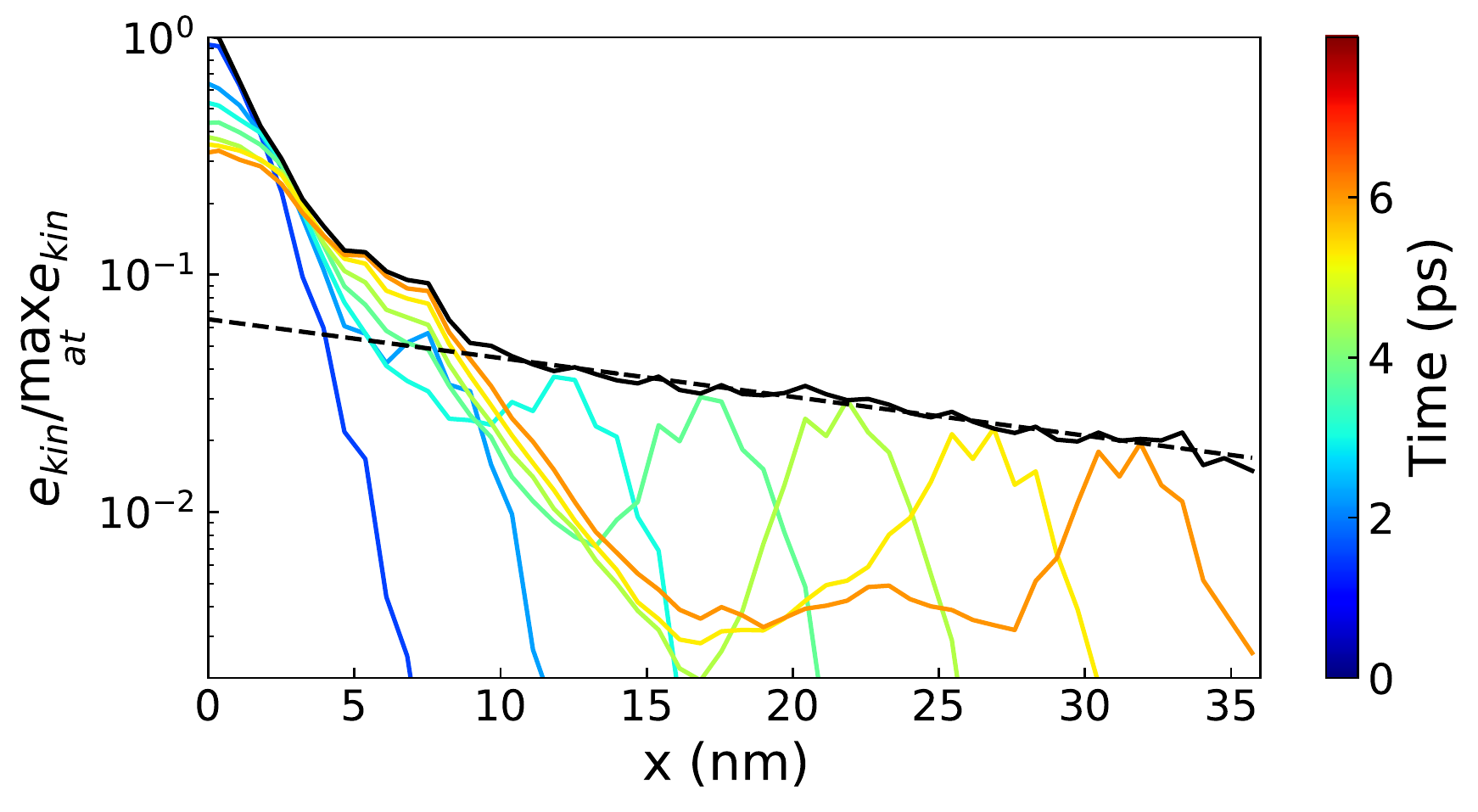}
	\caption{\label{fig:EnvEx} Representation of the propagation of an 8 THz longitudinal impulsion in the NW-M, with the energy distribution at different time steps (colored lines), the envelope defined as the maximal value of the kinetic energy at a given position (black solid line) and the exponential fit used to compute the MFP (dashed black line).}
\end{figure} 

The MFPs of the propagating modes are estimated from the decay rate of the envelope of the WP as a function of the distance to the excited slice. The envelope of the WP is defined as the maximum value of the kinetic energy at each point along the propagation path (see Figure~\ref{fig:EnvEx}). The envelope of a WP traveling ballistically follows a Beer-Lambert law (exponential decrease)~\cite{Beltukov2016}. Due to the presence of the NIs, the envelope may contain plateaus and sharp decreases, thus to get a meaningful value of the exponential decay fit, the portion on which the least square fit is made has to be chosen appropriately. Moreover, as visible in Figure~\ref{fig:EnvEx} in the vicinity of the excited slice a diffusive part is visible, this part is not included in the MFP computation. The propagation taking place in both $x$ positive and $x$ negative directions the final value of the decay rate is the average between the two.

At high frequencies the exponential decay can be ill-defined. This is the case for configurations without NIs above 12~THz for the longitudinal polarization, and for frequencies above 7~THz for the transverse polarization. In those cases, the decay does not follow an exponential attenuation. The penetration length is then used instead of the MFP. The penetration length is defined as the distance to the excitation point for which the energy as been divided by $e$~\cite{Beltukov2018}. This corresponds to the MFP in the case of a perfect exponential attenuation.

The energy diffusivity is estimated with the method described by Beltukov \textit{et al.}~\cite{Beltukov2013}. This is done after a random force excitation, to cancel the propagative (coherent) part. The average square distance to the diffusion front for each frequency is computed as: 

\begin{equation}
R^2(t)=\frac{1}{E_{tot}}\sum^N_{i=0}x_i^2 E_i,
\label{eq:R2}
\end{equation}

with $N$ the number of slices, $i$ the slice index, $x_i$ the distance to the excitation, $E_i$ the kinetic energy of the $i^{th}$ slice. The diffusivity is linked to the time dependence of $R^2$ by the equation of one dimensional diffusion,

\begin{equation}
R^2(t)=2D(\nu)t.
\label{eq:Dif}
\end{equation}

In each case $D(\nu)$ is computed through a least square fit of $R^2(t)$.
\subsection{Lifetime Estimation and Temperature Effect} 

The computation of the thermal conductivity through Equation~\ref{eq:kp} relies on the estimation of the phonon lifetime as a function of frequency and polarization. The lifetime is considered to be limited by two phenomena: interfaces or defect scattering, and the phonon-phonon scattering. The former is assumed to be geometry dependent only and is estimated thanks to the MFP and the group velocity:

\begin{equation}
\tau^{-1}_{geom} =\frac{v_{\eta}(\nu)}{\Lambda_{\eta}(\nu)}
\label{eq:TauGeom}
\end{equation}

with $\Lambda_\eta(\nu)$ the MFP at frequency $\nu$.

The wave-packet propagation simulation taking place at 0~K, the reduction of lifetime due to anhamonicity is underestimated. To compensate for this, a lifetime due to phonon-phonon interactions is introduced. This lifetime is estimated with the empirical relation described in the Callaway model as a function of temperature and frequency~\cite{Callaway1959}.

\begin{equation}
\tau^{-1}_{ph-ph} =P(2\pi\nu)^2T\exp(-C_U/T) 
\label{eq:TauUmklapp}
\end{equation} 

with $P$ and $C_U$ empirical scattering parameters, the used values are the one of crystalline bulk silicon found in the publication of Yang \textit{et al.}~\cite{Yang2013}.

The global lifetime used in Equation~\ref{eq:kp} is then estimated using Matthiesen summation rule:

\begin{equation}
\tau^{-1}=\tau^{-1}_{geom}+\tau^{-1}_{ph-ph}.
\label{eq:TauMatt}
\end{equation}

\subsection{Group Velocity through the Dynamical Structure Factor}
\label{sec:DSF}

The dynamical structure factor (DSF) is a spatial and temporal Fourier transform of the atomic displacements used to characterize the vibrational properties of a system. This is very similar to what can be measured by X-ray or by neutrons scattering experiments~\cite{Yip1980}. It is defined as: 

\begin{equation}
S(\mathbf{q},\omega)=\frac{2}{NT}\left|\sum_i^{N_{at}} \exp(-i\mathbf{q}\cdot \mathbf{r_i})\int_{0}^{\tau}\mathbf{u_i}(r_i,t)\mathbf{m}_\eta exp(i\omega t)dt \right|^2
\label{eq:DSF}
\end{equation}

with $\mathbf{q}$ the wave vector, $\mathbf{u_i}$ and $\mathbf{r_i}$ the displacement and position of the $i^{th}$ atom, $\mathbf{m}_\eta$ the polarization vector (parallel or perpendicular to $\mathbf{q}$), $T$ the temperature and $N$ the total number of atoms~\cite{Damart2015}. 

The resolution of the wave vector is given by $2\pi/L$ with $L$ the length of the simulation box in the direction of the wave vector. The direction of the vector $\mathbf{q}$ can be chosen arbitrarily to match the different direction in the reciprocal lattice space (here that of c-Si).

\begin{figure}[h] 
	\centering
	\includegraphics[width=.8\textwidth]{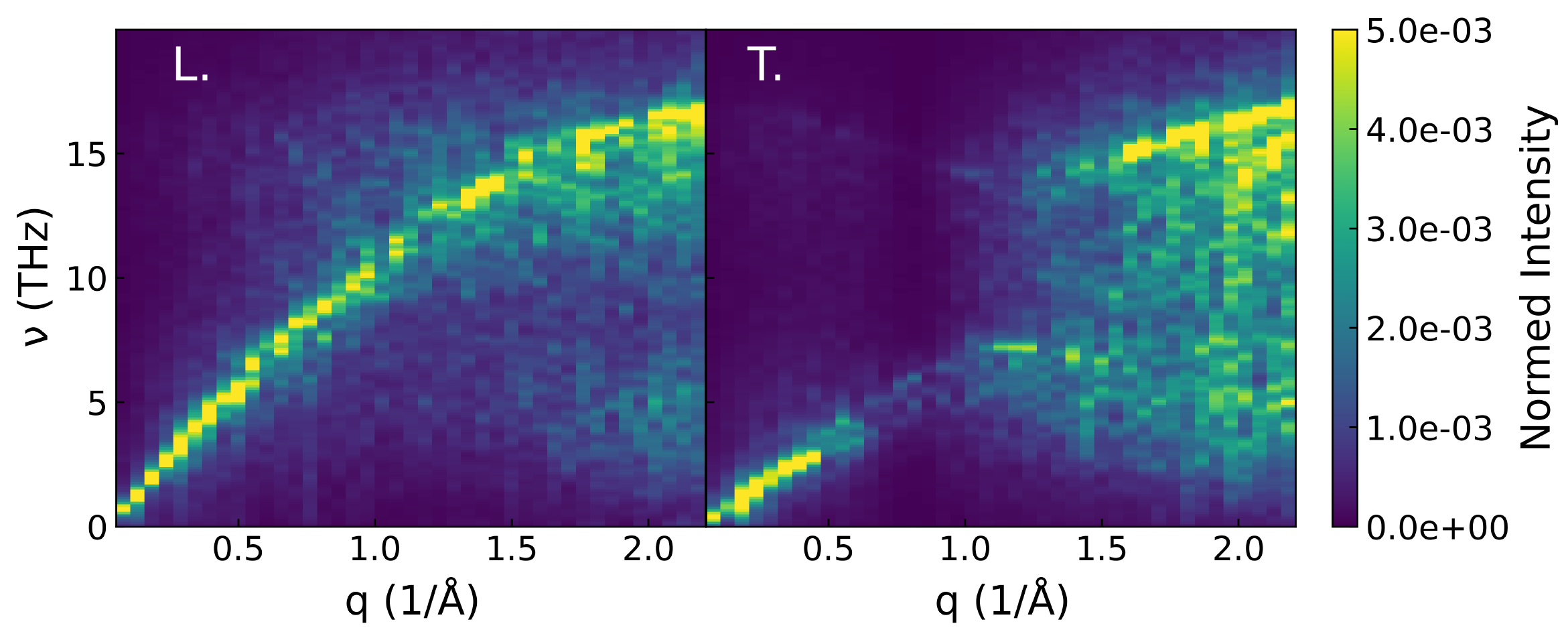}
	\caption{\label{fig:DynStructFact}Dynamical structure factor computed through Equation~\ref{eq:DSF} for the spherical NI in the $\Gamma X$ direction for the longitudinal (left panel) and transverse polarization (right panel).}
\end{figure}

The atomic trajectories used for the computation of $S(\mathbf{q},\omega)$ are obtained in the following manner: the sample is heated at 100~K and equilibrated at this temperature for \num{5e-3}~ns using a Nos\'{e}-Hoover thermostat. After this, the atomic trajectories are recorded during a \num{1e-2}~ns long constant energy simulation. An example of DSF is displayed in Figure~\ref{fig:DynStructFact}.

From the DSF the phononic dispersion curves can be obtained. First, the DSF is filtered through a convolution with a typical energy resolution curve of line-width 1.35 meV (as suggested by Tlili et \textit{al.}~\cite{Tlili2019}). Then, for a given wave-vector direction, the dispersion is estimated from the frequency for which $S(\mathbf{q},\omega)$ has the highest value for each wave vector within the acoustic phonons frequency range. This dispersion is finally fitted to a sine function allowing the analytical derivation of the group velocity as a function of frequency. The expression of the group velocity contains an $\arcsin$ function, thus when the frequency is outside of the definition domain it becomes ill-defined and will be considered nil. To get the appropriate dispersion, $\mathbf{q}$ is chosen as the propagation direction of the WP. This corresponds to the $\langle 100 \rangle$ crystalline orientation in the direct space or to $\Gamma X$ in the reciprocal space. An alternative method to estimate both the dispersion relation and the lifetime from the DSF is discussed in the appendix~\ref{app:DHO}.

\subsection{Vibrational Density of States}
The vibrational density of states (VDOS) of the different configurations is evaluated with the Fourier transform of the velocity auto-correlation function (VACF)~\cite{Dove1993}. Before computing the VACF the system is equilibrated at 50~\si{\K} for 0.1~ns with a Nos\'{e}-Hoover thermostat. The VACF averaged over all the atoms is then recorded over the next 0.1~ns without thermostat. To obtain the final VDOS, the Fourier transform of the VACF is filtered using a Savitzky-Golay polynomial filter~\cite{Savitzky1964}. 

Additionally, the VDOS of the amorphous Si was computed using the dynamical matrix~\cite{Dove1993} on a smaller sample. The square roots of the eigenvalues of this matrix give the eigenfrequency of the system. By distinguishing the modes keeping the volume of the Voronoi cell around each atom and those who do not, the transverse and longitudinal modes can be distinguished~\cite{Beltukov2015}. The VDOS is then approximated by series of Chebychev polynomials~\cite{Weiss2006}. The dynamical matrix was computed for a cubic cell of side 4~nm with periodic boundary conditions containing 3159 atoms and the Voronoi cells determined thanks to the \textit{Voro++} open-source software~\cite{Rycroft2007}.

\section{Results}
\subsection{Ballisticity through Wave-Packet Simulations}

A qualitative analysis of the time evolution of the kinetic energy distribution can give physical insights of the impact of nanostructuration on energy propagation. Table~\ref{tab:LFL} shows the atomic kinetic energy on a cross-section for the different configurations after a 2~THz impulsion. The impulsion is made in the middle of the system and propagates in both the negative and positive $x$ directions. The two directions being symmetric only one direction ($x$ positive) is represented. The first half of the table corresponds to longitudinal polarization. The main observation for most configurations, is that most of the energy travels through the sample as a plane wave. The NIs do not strongly affect the propagation at this frequency: the WP travels through the nanocomposites and the a-Si similarly. However, there is still some scattering visible through the small spots of high energy concentration after the passage of the WP. These spots are mainly located in a-Si and at the interfaces between NIs and matrix.
For the porous configuration a plane wave is also visible, albeit, its intensity is strongly reduced by the time it reaches the end of the simulation box. But more importantly, most of the energy stays in the center and slowly diffuses through the sample.

\begin{table}[h] 
	\caption{\label{tab:LFL}Cross-sectional view of a WP going through the different systems after a longitudinal excitation at 2~THz (first part of the table) or a transverse excitation at the same frequency (second part of the table) every .9~ps. The first line represents the geometry of the cross-sections at the middle of simulation box with inclusions in yellow and matrix in dark grey. The color scale going from 0 (blue) to \SI{3e-9}{\eV} (dark red) gives the atomic kinetic energy.}
	\centering
	\begin{tabular}{cccccc}
		\toprule
		\textbf{Amorphous}&	\textbf{Pore}&\textbf{Sphere}&\textbf{SC}&\textbf{STC}&\textbf{NW-M}\\
		\midrule
		\includegraphics[width=.135\textwidth]{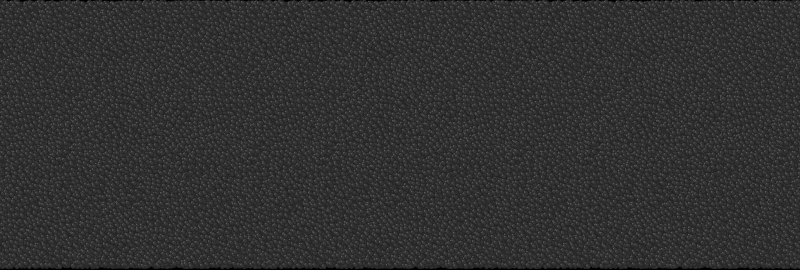}&\includegraphics[width=.135\textwidth]{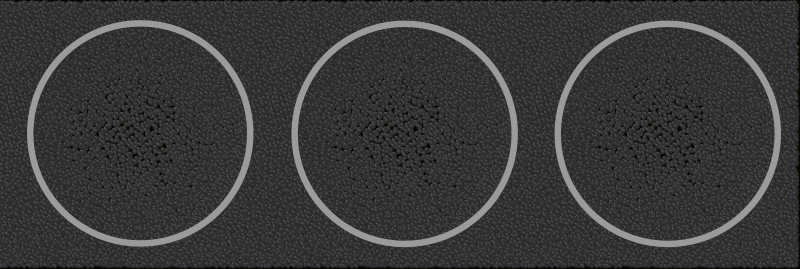}&\includegraphics[width=.135\textwidth]{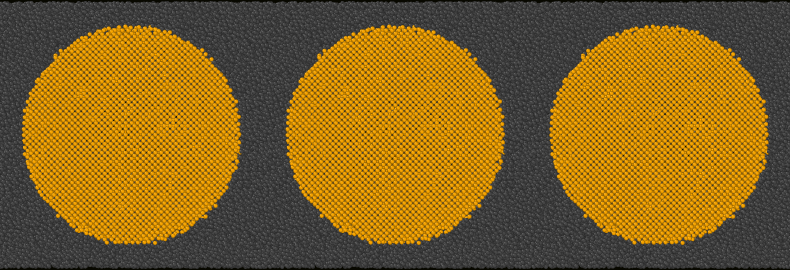}&\includegraphics[width=.135\textwidth]{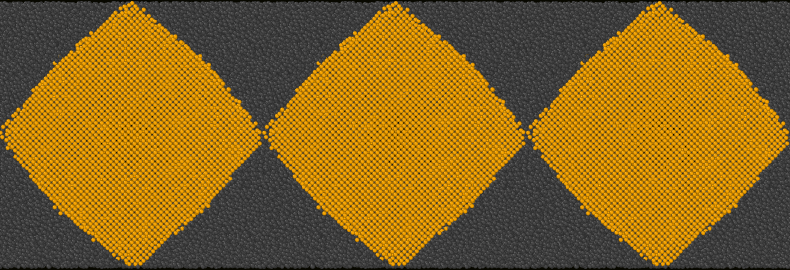}&\includegraphics[width=.135\textwidth]{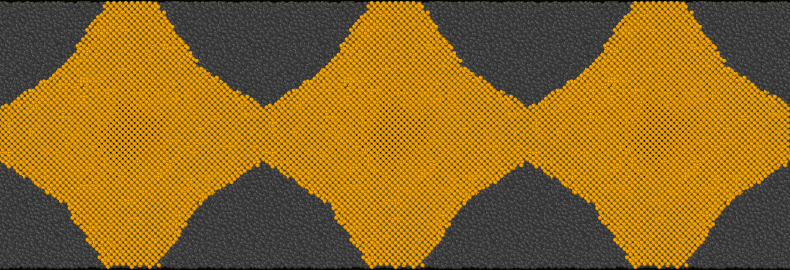}&\includegraphics[width=.135\textwidth]{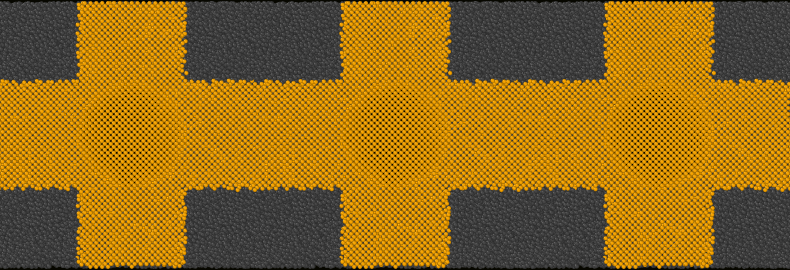}\\
		\midrule
		\includegraphics[width=.05\textwidth]{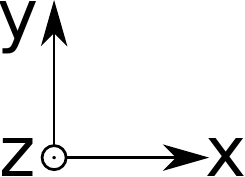}&\textbf{Longitudinal}&2~THz&\multicolumn{3}{c}{\includegraphics[width=.2\textwidth]{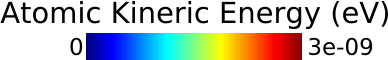}}\\
		\midrule
		\TabLLineLF{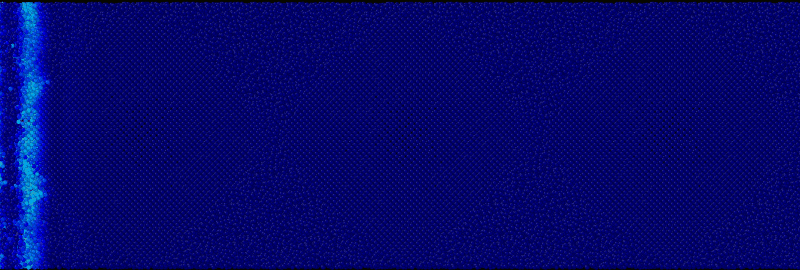}
		\TabLLineLF{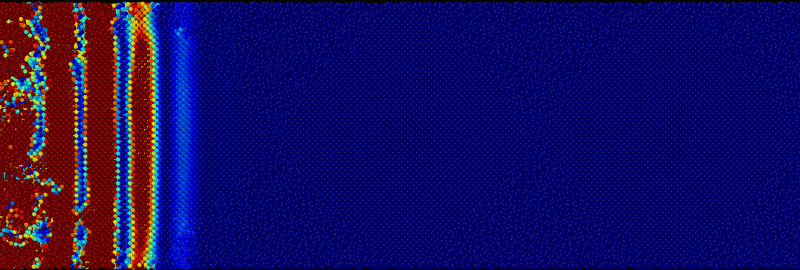}
		\TabLLineLF{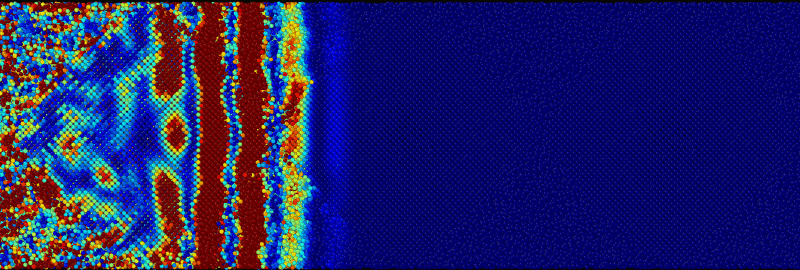}
		\TabLLineLF{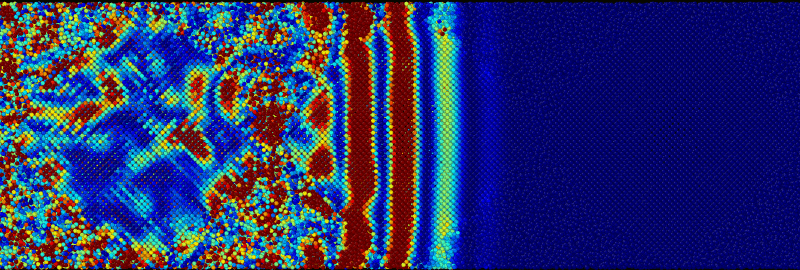}
		\TabLLineLF{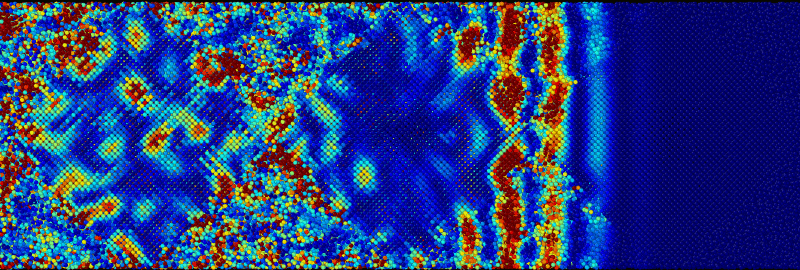}
		\TabLLineLF{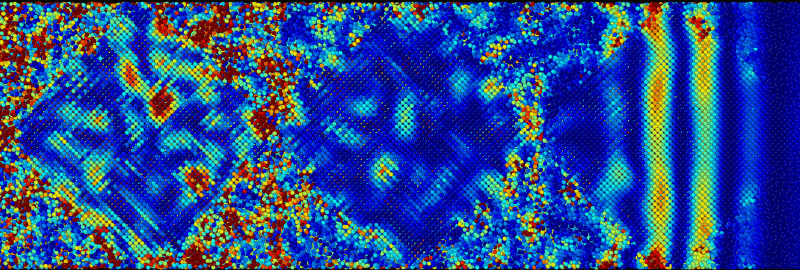}
		\TabLLineLF{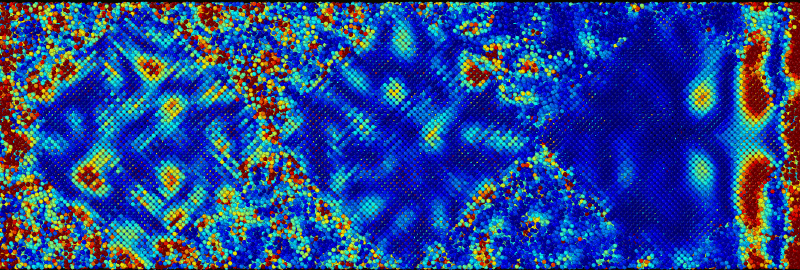}
		\midrule
		\includegraphics[width=.05\textwidth]{Coord}&\textbf{Transverse} &2~THz&\multicolumn{3}{c}{\includegraphics[width=.2\textwidth]{ColorBar}}\\
		\midrule
		\TabTLineLF{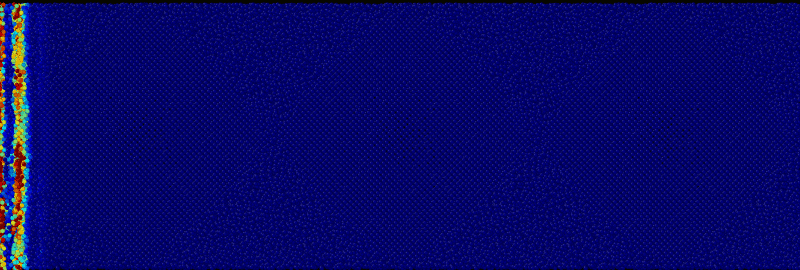}
		\TabTLineLF{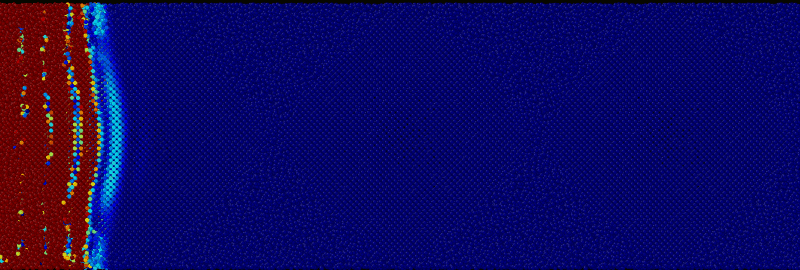}
		\TabTLineLF{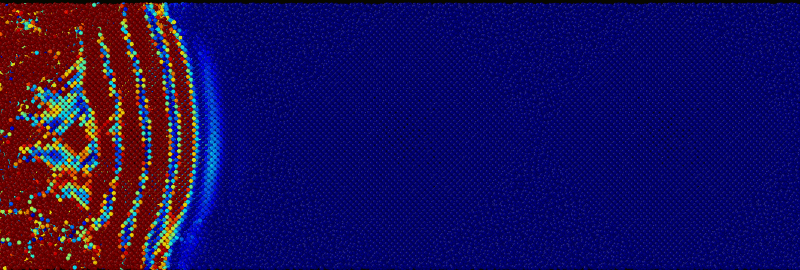}
		\TabTLineLF{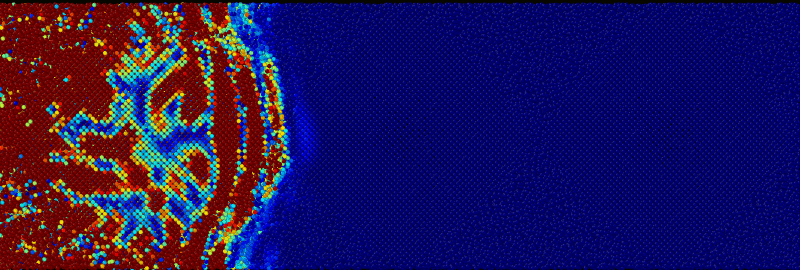}
		\TabTLineLF{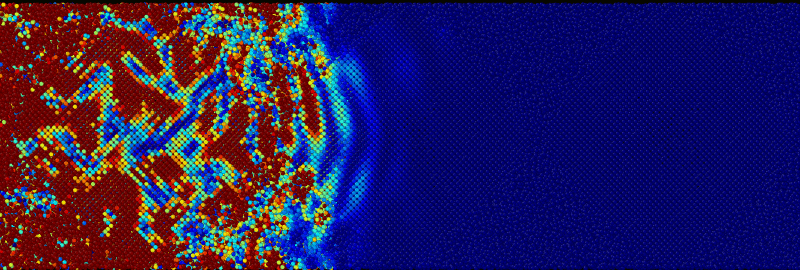}
		\TabTLineLF{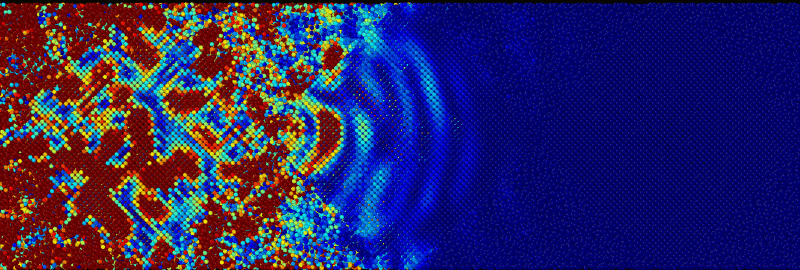}
		\TabTLineLF{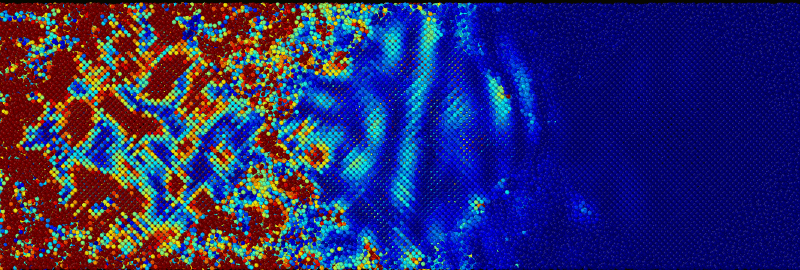}
		\bottomrule
	\end{tabular}
\end{table}

For the transverse waves at the same frequency (displayed in the second half of Table~\ref{tab:LFL}), the dispersion is more marked. The vertical red lines, characteristic of plane waves, can be distinguished in the first few images, but disappear before reaching the simulation box boundary. The waves are quickly scattered, and this even for bulk a-Si. In the configurations containing NIs, the vertical lines materializing the plane waves are distorted. This distortion of the wave-front is due to the WP traveling faster in the crystal than in the glass matrix.
The porous configuration is again, the configuration for which the scattering is the strongest.

To sum up about the low frequency WP propagation, one can observe that the shape of the NIs has no impact on either the longitudinal or transverse waves. The longitudinal plane-waves preserve their shape for both interconnected and not interconnected NIs, and the transverse waves are diffused quickly. The situation is quite different for the nanoporous amorphous silicon, for which the plane waves disappear rapidly for both polarizations. We stress the fact that an amorphous/crystalline nanocomposite could be "transparent" to low frequency longitudinal waves.

The behavior of the nanocomposites after a high frequency impulsion is displayed in Table~\ref{tab:HFL}. The first part contains the evolution of a longitudinal WP at 10~THz or two third of the maximum frequency for which a group velocity can be defined. It appears that there is no propagation in the amorphous matrix. For all configurations, the energy slowly spreads through the amorphous matrix.

However, on top of this diffusion, a propagative behavior limited to the crystal also appears. This is particularly noticeable in case of structural percolation. In this case, the wave packet takes an oval shape and travels through the structural percolation. In absence of percolation, the propagative part of the WP is scattered at the first crystalline/amorphous interface. 

\begin{table}[h]
	\caption{\label{tab:HFL}Cross-sectional view of a WP going through the different systems after a longitudinal excitation at 10~THz (first part of the table) or a transverse excitation at 4~THz (second part of the table) every .9~ps. The first line represents the geometry of the cross-sections at the middle of simulation box with inclusions in yellow and matrix in dark grey. The color scale going from 0 (blue) to \SI{3e-9}{\eV} (dark red) gives the atomic kinetic energy.}
	\centering
	\begin{tabular}{cccccc}
		\toprule
		\textbf{a-Si}&\textbf{Pore}&\textbf{Sphere}&\textbf{SC}&\textbf{STC}&\textbf{NW-M}\\
		\midrule
		\includegraphics[width=.135\textwidth]{3Am}&\includegraphics[width=.135\textwidth]{3Pore1}&\includegraphics[width=.135\textwidth]{3Sphere}&\includegraphics[width=.135\textwidth]{3Cone}&\includegraphics[width=.135\textwidth]{3ConeTronque}&\includegraphics[width=.135\textwidth]{3NanoFil}\\
		\midrule
		\includegraphics[width=.05\textwidth]{Coord}&\textbf{Longitudinal}& 10~THz&\multicolumn{3}{c}{\includegraphics[width=.2\textwidth]{ColorBar}}\\
		\midrule
		\TabLLineHF{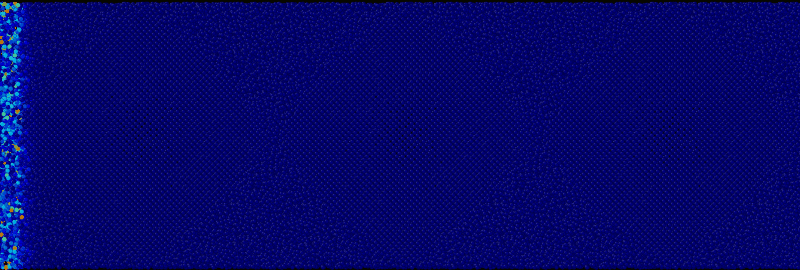}
		\TabLLineHF{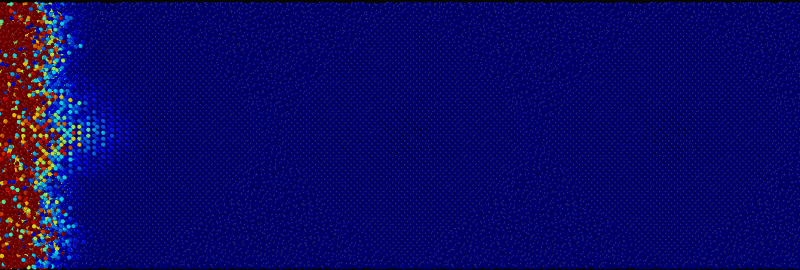}
		\TabLLineHF{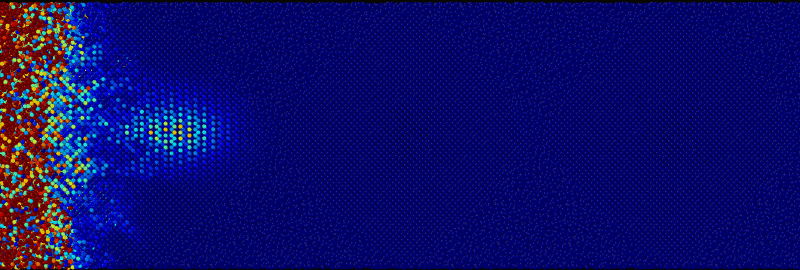}
		\TabLLineHF{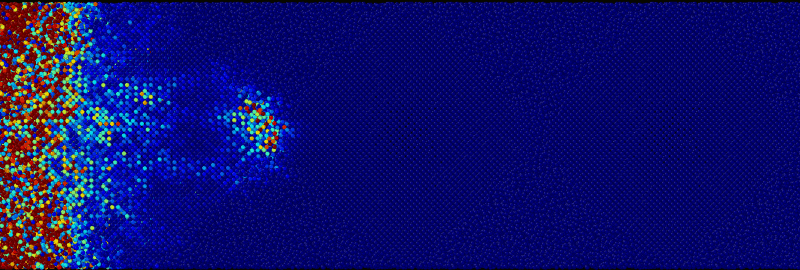}
		\TabLLineHF{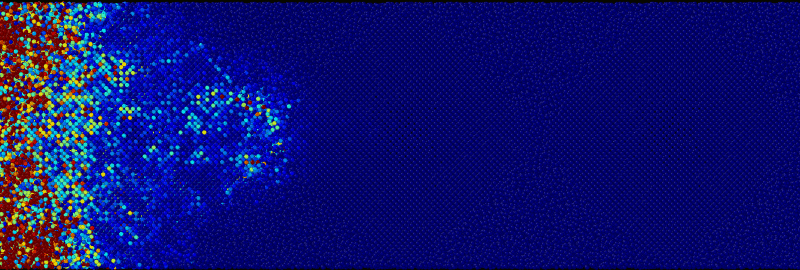}
		\TabLLineHF{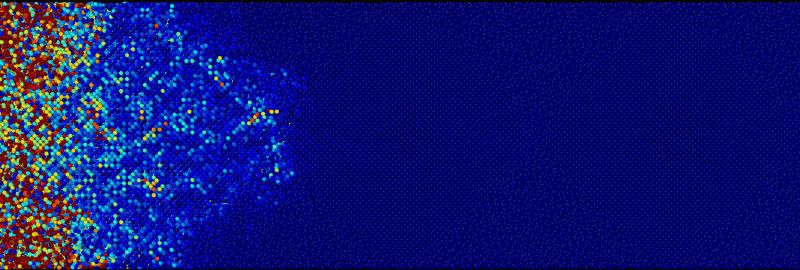}
		\TabLLineHF{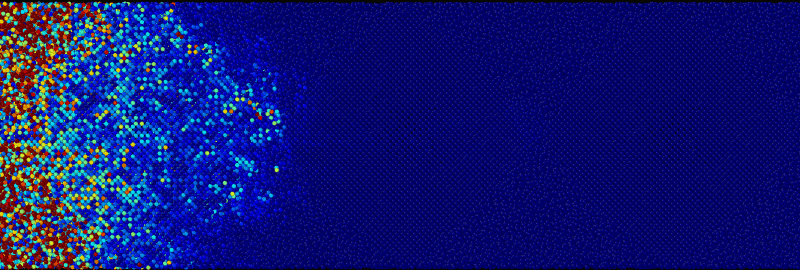}
		\midrule
		\includegraphics[width=.05\textwidth]{Coord}&	\textbf{Transverse}& 4~THz&\multicolumn{3}{c}{\includegraphics[width=.2\textwidth]{ColorBar}}\\
		\midrule
		\TabTLineHF{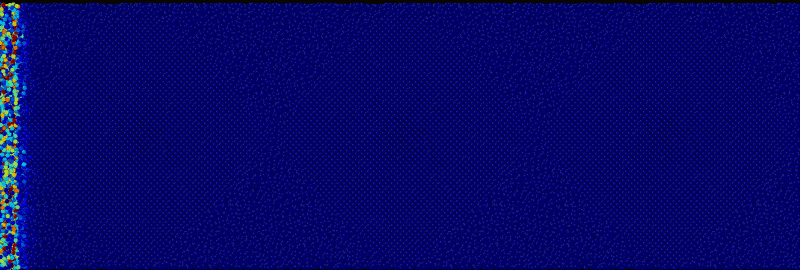}
		\TabTLineHF{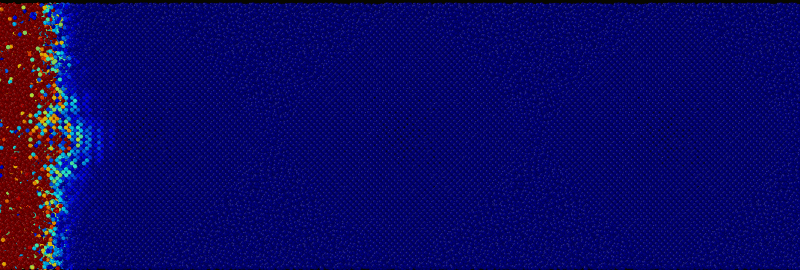}
		\TabTLineHF{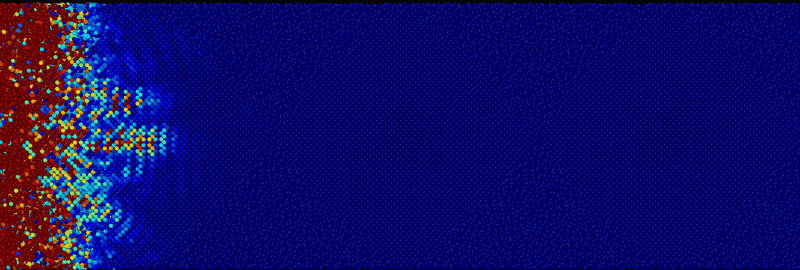}
		\TabTLineHF{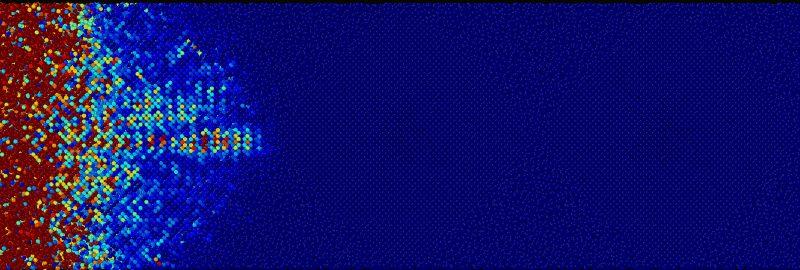}
		\TabTLineHF{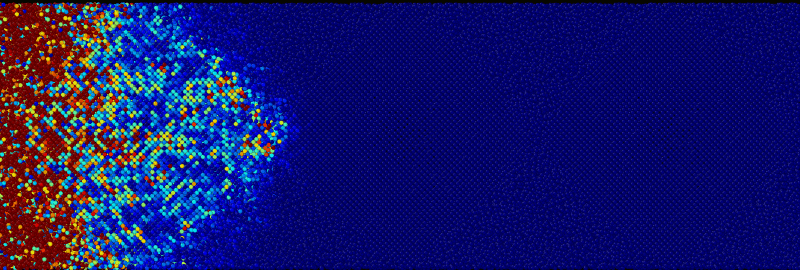}
		\TabTLineHF{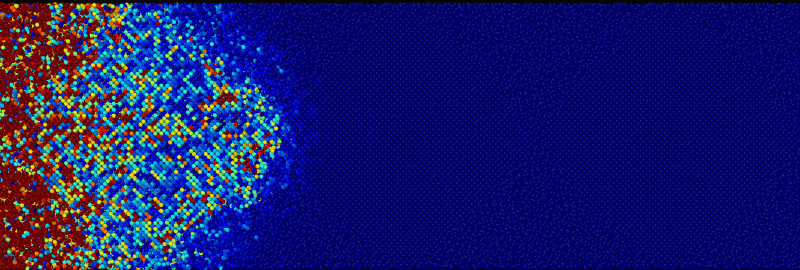}
		\TabTLineHF{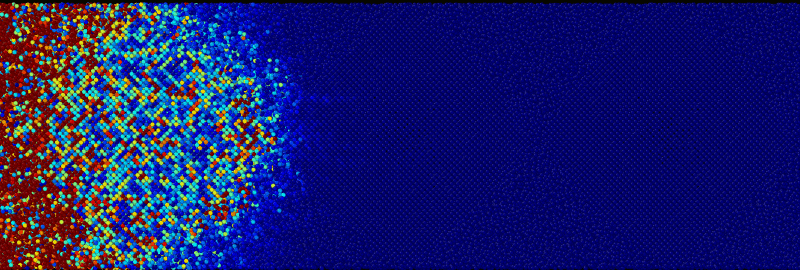}
		\bottomrule
	\end{tabular}
\end{table}
For the transverse polarization, the selected frequency is 4 THz. As for the longitudinal polarization, this frequency corresponds approximately to two third of the frequency for which the group velocity becomes ill-defined (see Figure \ref{fig:KappaProp}). The behavior is very similar to the longitudinal polarization: there is ballistic transport limited to the structural percolation region and a diffusive transport acting on a slower timescale. This diffusive behavior is visible close to the border, where the impulsion is made. However, in this case the both the crystalline NIs and the amorphous matrix participate in diffusive energy transport.

To sum up concerning the WP propagation at high frequencies, there is a clear differentiation of the crystalline and amorphous phases. There is no propagation in the amorphous phase. Ballistic propagation through the sample is only possible through the structural percolation. We also observe that there is no backscattering or important deviation of energy in the perpendicular branches of the inclusions.

\subsection{Diffusive and Propagative Contributions to the Thermal Conductivity}

\begin{figure}[h]
	\centering
	\includegraphics[width=.7\textwidth]{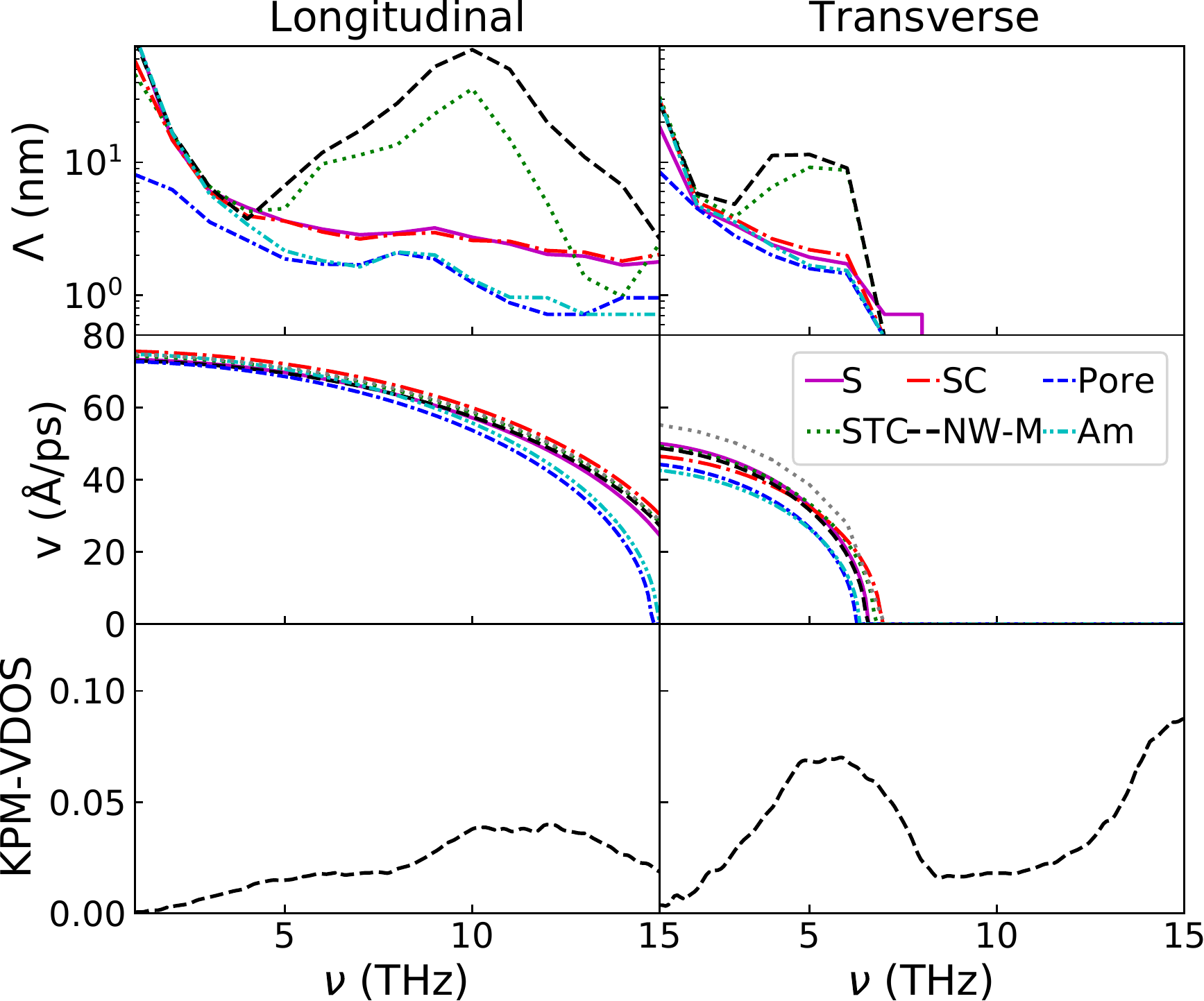}
	\caption{\label{fig:KappaProp}From right to left: first row longitudinal and transverse mean free path, second longitudinal and transverse group velocity for the studied configurations, third row longitudinal and transverse VDOS. Additionally, the group velocity computed for a fully crystalline sample is displayed with a dotted gray line.}
\end{figure}

As described in section~\ref{sec:methode}, information extracted from the WP simulations can be used to estimate the thermal conductivity. First, the different components of the propagative contribution to the thermal conductivity ($\kappa_P$) are displayed in Figure~\ref{fig:KappaProp}. In the top left panel the MFPs of the longitudinal WP for the different configurations are displayed as a function of the frequency. The MFP is estimated through the penetration length for frequencies above 12 THz for the amorphous and porous configurations. These curves confirm what was visible in Table \ref{tab:LFL}: the MFP is high at low frequencies for all configurations. Below 5~THz, the MFP of the non-porous configuration are very similar. Only the pores decrease the MFP at low frequencies. At higher frequencies, the configurations without structural percolation have a low MFP. This contrasts with the configurations with structural percolation, for which the MFP rises between 5 and 10~THz and decreases strongly after that. The MFP for those configurations, around its maximum between 8 and 12~THz, is almost one order of magnitude higher than without percolation. Moreover, the interconnection degree has an influence. The MFP is higher for the NW-M than for the STC. It is also noticeable, that the porous and fully amorphous configurations have a small MFP peak around 8~THz, this peak was already observed for a-Si by Beltukov \textit{et al.}~\cite{Beltukov2018}. It has been associated to the decreased number of transverse modes available for coupling at this frequency.

For the transverse polarization, in the top right panel of Figure~\ref{fig:KappaProp} the behavior is similar. Above 7~THz the MFP is substituted by the penetration length for all the configurations in order to avoid artifacts caused by a strong scattering. As for the longitudinal polarization, the MFPs of the configurations with structural percolation have a maximum. In this case, the maximum lays between 5-6~THz. Without structural percolation, the MFP decreases as the frequency increases. Again, similarly to the longitudinal polarization, most configurations share a very similar MFP at 1~THz, the only exception being the porous configuration, that has a lower MFP.

The group velocities for the longitudinal and transverse polarizations are displayed in the second row of Figure~\ref{fig:KappaProp}. All the configurations share a very similar group velocity. This is especially true for the longitudinal polarization at low frequencies (below 5~THz). At higher frequencies, the group velocities for the amorphous and porous configurations are lower than the group velocities of the others. The $v$ of the configurations containing NIs are very similar to the $v$ of c-Si. For the transverse polarization, there is also a group velocity difference, although spanning over the whole spectrum. For this polarization the $v$ of the nanocomposites containing NIs lays in between those of c-Si and a-Si. Finally, the transverse polarization has a nil velocity for frequencies higher than 7~THz.

The third row from the top contains the VDOS attributed to the longitudinal and transverse polarizations for the computation of $\kappa_P$. For this application, the transverse and longitudinal VDOS are computed via the KPM method \cite{Beltukov2015} on an a-Si sample. This allows for a good approximation in the 0-12~THz frequency range (see appendix~\ref{app:VDOSDSF} for more detail). On these graphs it can be noted that the maximum of VDOS at 10~THz for the longitudinal polarization and at 5~THz for the transverse also correspond to MFP maxima. Due to the higher lifetime conjunct with a high VDOS, these modes will contribute significantly to $\kappa_{P}$.

\begin{figure}[h]
	\centering
	\includegraphics[width=.5\textwidth]{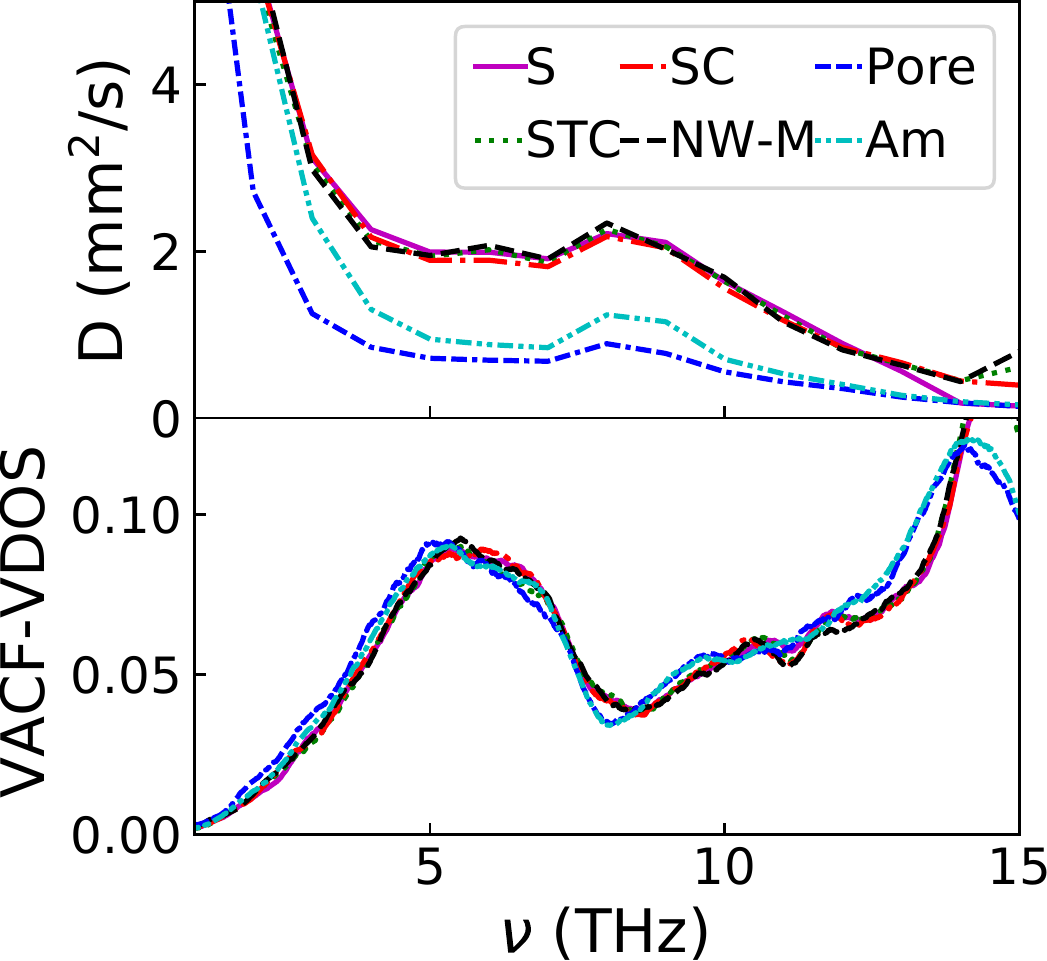}
	\caption{\label{fig:KappaDiff} Diffusivity and VDOS as a function of frequency for the studied configurations.}
\end{figure}

The different terms contributing to $\kappa_D$ are shown in Figure~\ref{fig:KappaDiff}. The top panel corresponds to the diffusivity computed with Equation~\ref{eq:Dif}. Two main observations can be made: firstly all the configurations containing a crystalline phase share a very similar diffusivity across the whole spectrum, secondly only the porous configuration induces a reduction of diffusivity with respect to the amorphous sample. The addition of NIs increases the diffusivity.
Additionally, a small peak at 8~THz is visible for all cases, this peak corresponds to the end of the transverse phonon dispersion curve and was already observed by Allen and Feldman~\cite{Allen1999}.
The VDOS computed through the VACF for the different configurations are displayed in the bottom panel of Figure~\ref{fig:KappaDiff}. All the VDOS are very similar up to 14~THz. At higher frequencies the configurations containing NIs and the others show differences. The VDOS of a-Si starts to decrease from 14~THz while the others continue to increase. However, this difference has little effect on the $\kappa_{D}$ given that the diffusivity is very low at those frequencies.

\begin{figure}[h]
	\centering
	\includegraphics[width=\textwidth]{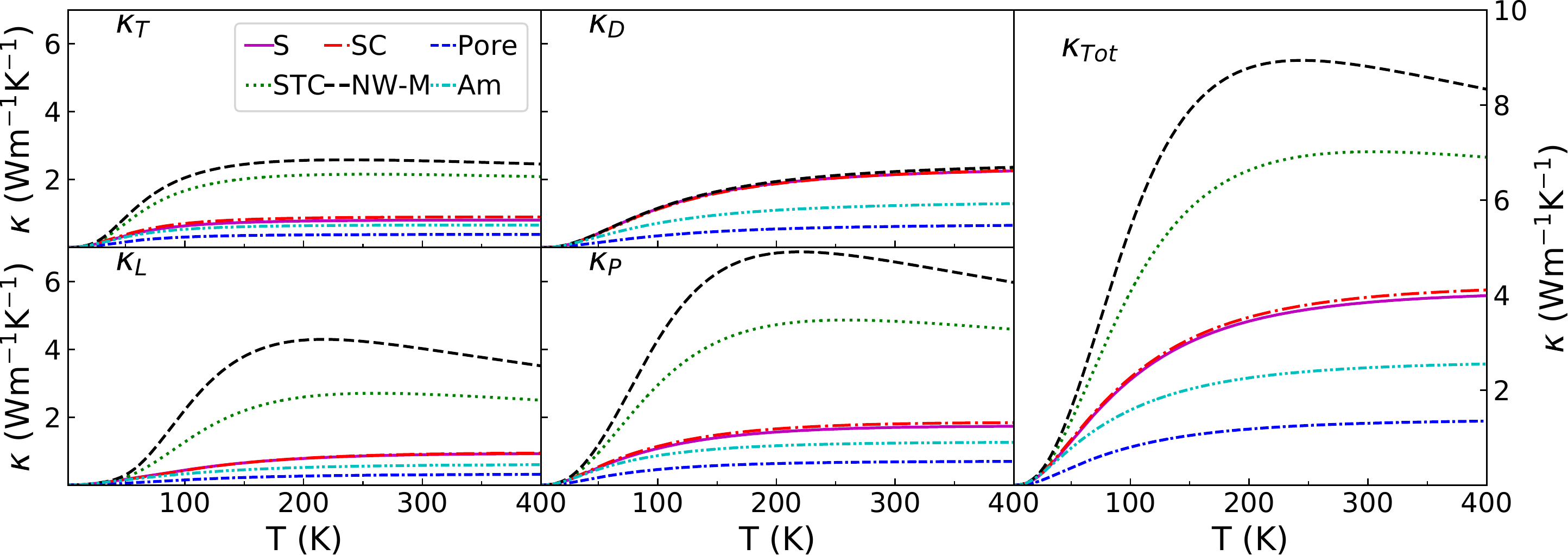}
	\caption{\label{fig:KappaDet} Different contributions to the thermal conductivity, the first column contains the contributions of the longitudinal phonons $\kappa_L$ (bottom) and the transverse phonons $\kappa_T$ (top), the second column the diffusive contribution $\kappa_D$ (top) and the overall propagative contribution $\kappa_P=\kappa_L+\kappa_T$ (bottom), the last columns contains the sum of the different contribution $\kappa_{Tot}=\kappa_P+\kappa_D$.}
	
\end{figure}

The different terms displayed in Figures~\ref{fig:KappaProp} and \ref{fig:KappaDiff} are used to compute $\kappa_{P}$ and $\kappa_{D}$. The results for temperatures between 10 and 400~K are displayed in Figure~\ref{fig:KappaDet}. The first column contains the transverse and longitudinal propagative contribution. It confirms that the structural percolation induces a marked increase of the propagative contribution, the STC and NW-M have a larger $\kappa_T$ and $\kappa_L$. However, for the diffusive contribution in the top row of the central column, no distinction between the configurations containing NIs can be made. Only the pores seem to decrease the diffusive contribution below amorphous values. The propagative contribution, for both polarizations increases with the degree of interconnection. When looking at the propagative contribution as a function of the temperature, it appears that $\kappa_L$ increases at higher temperature than $\kappa_T$. This is linked to the MFP peak at 10~THz and to $C(T,\omega)$ that limits the impact of high frequencies at low temperature. This important high frequency contribution also results in a maximum of $\kappa_L$ around 200~K for the NW-M. This is due to the empirically added phonon-phonon term (Equation~\ref{eq:TauUmklapp}) that reduces the contribution of high frequency phonons as the temperature rises. 
The different contributions (propagative and diffusive) can be compared in the central panel. The diffusive and propagative contributions for the non percolating configurations have similar values at 300~K. 

The sum of the different contributions, $\kappa_{Tot}$, is displayed in the last column of Figure~\ref{fig:KappaDet}. At all temperatures, the same order of $\kappa_{Tot}$ is preserved. This order is, from the highest to the lowest thermal conductivity, the NW-M, the STC, the SC and the S at very similar value, the amorphous, and finally the porous configuration. The maximum observed for $\kappa_L$ of the NW-M is still visible on the sum and happens at 244~K. Such a maxima in $\kappa$ has already been predicted for SiC NWs using a similar method~\cite{Chantrenne2012} but contrasts with experimental results on Si NW~\cite{Li2003b}.

\begin{figure}[h]
	\centering
	\includegraphics[width=.5\textwidth]{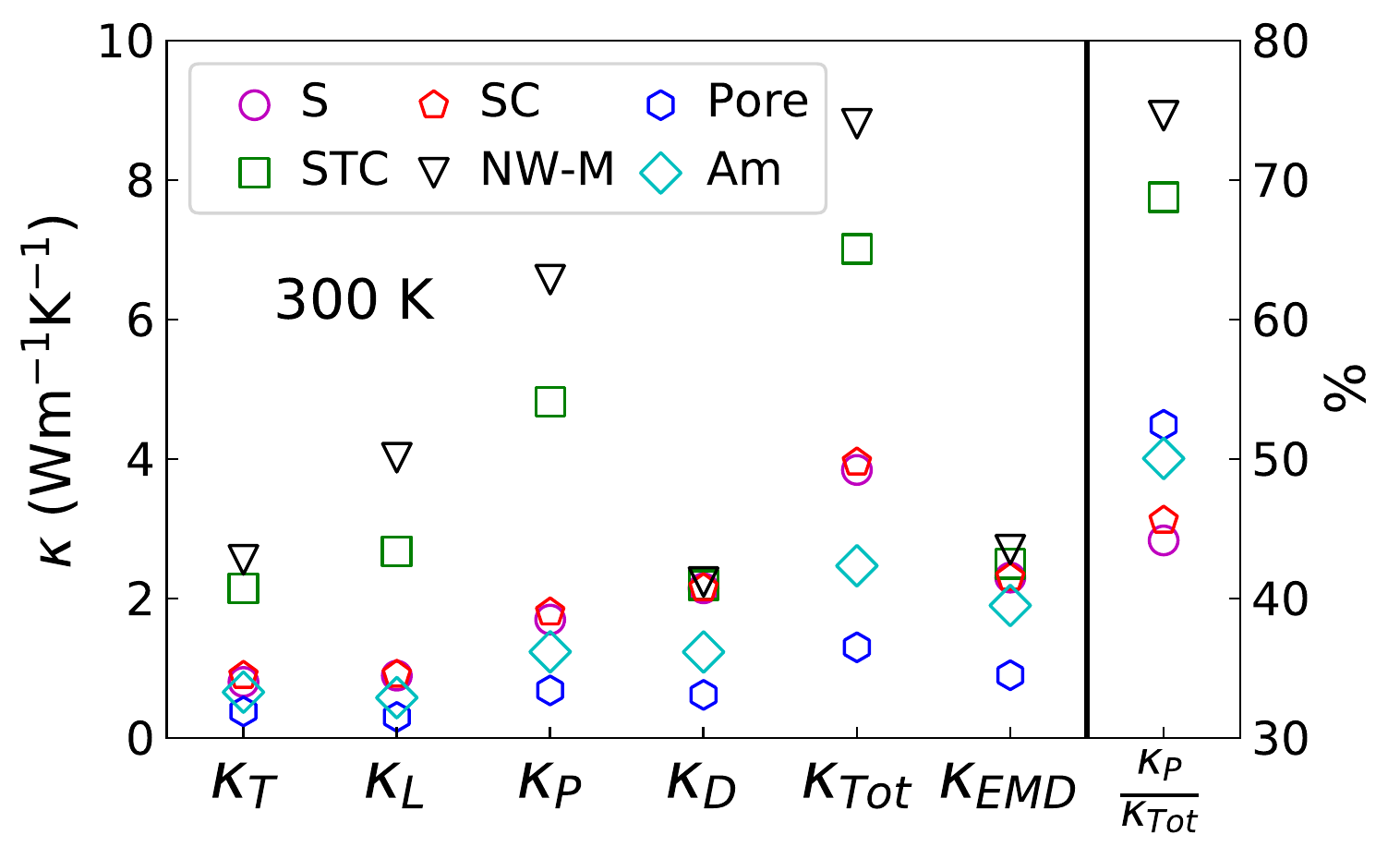}
	\caption{\label{fig:KappaComp300}Thermal conductivities at 300 K, decomposed through equations \ref{eq:kp} and \ref{eq:kd} and computed with EMD. The lines are a guide for the eye, The proportional contribution to the thermal conductivity of the propagative mode is also represented in the right part of the figure (right axis). }
\end{figure}

The different contributions to the thermal conductivity at 300~K are also shown in Figure \ref{fig:KappaComp300}. With this representation it appears clearly that: the structural percolation increases $\kappa_{P}$ and does not affect $\kappa_{D}$. As a result the propagative part represent up to 75\% of $\kappa$ for those nanocomposites. 
This grao also shows that the addition of non-percolating NIs in an amorphous matrix increases more the diffusive transport than the propagative transport. For the S and SC configuration the diffusive transport is dominant. Finally, it appears that in spite of the overestimation of the thermal conductivity of nanocomposite containing NIs by the kinetic method compared to the results of EMD, the hierarchy in the different structures is preserved. 

To sum up briefly on the results obtained with the kinetic theory, it appears that as predicted previously the addition of NIs increases $\kappa$ above bulk a-Si values~\cite{Tlili2019}. This is due to the fact that the NIs are crystalline. This is particularly visible in the case of structural percolation, where the MFP peak at high frequencies is concomitant to a VDOS peak, results in a large increase of $\kappa_P$. This increase occurs mainly at high temperatures (see Figure \ref{fig:KappaDet}) due to the temperature dependent frequency weighting of $C(\nu,T)$ (see Equation \ref{eq:C}). As high frequencies at high temperatures are also more impacted by the phonon-phonon term (Equation~\ref{eq:TauUmklapp}), a maximum of $\kappa_{Tot}(T)$ appears for the NW-M. This maximum contrasts with experimental results for single nanowires of diameter similar to the NW constituting the NW-M. For these single NWs, no maximum of the thermal conductivity has been observed as a function of temperature~\cite{Li2003b}. This is a first sign that the propagative contribution, the only which can cause the apparition of a maximum of $\kappa$, is overestimated by our implementation of kinetic theory. The evolution of $\kappa$ with temperature is worth commenting: between 10 and 100 K all the configurations seem to follow the unusual $T^2$ power law as was observed experimentally below 1~K \cite{Zeller1971}, but it can certainly not be attributed here to double well potential effect since anharmonicity is not taken into account in our simulations in this temperature range. Moreover, $\kappa_{P}$ dominates at these temperatures, contrasting with the predictions of Cahill \textit{et al.} \cite{Cahill1988}.
\subsection{Global Estimation of the Thermal Conductivity}

\begin{figure}[h]
	\centering
	\includegraphics[width=.5\textwidth]{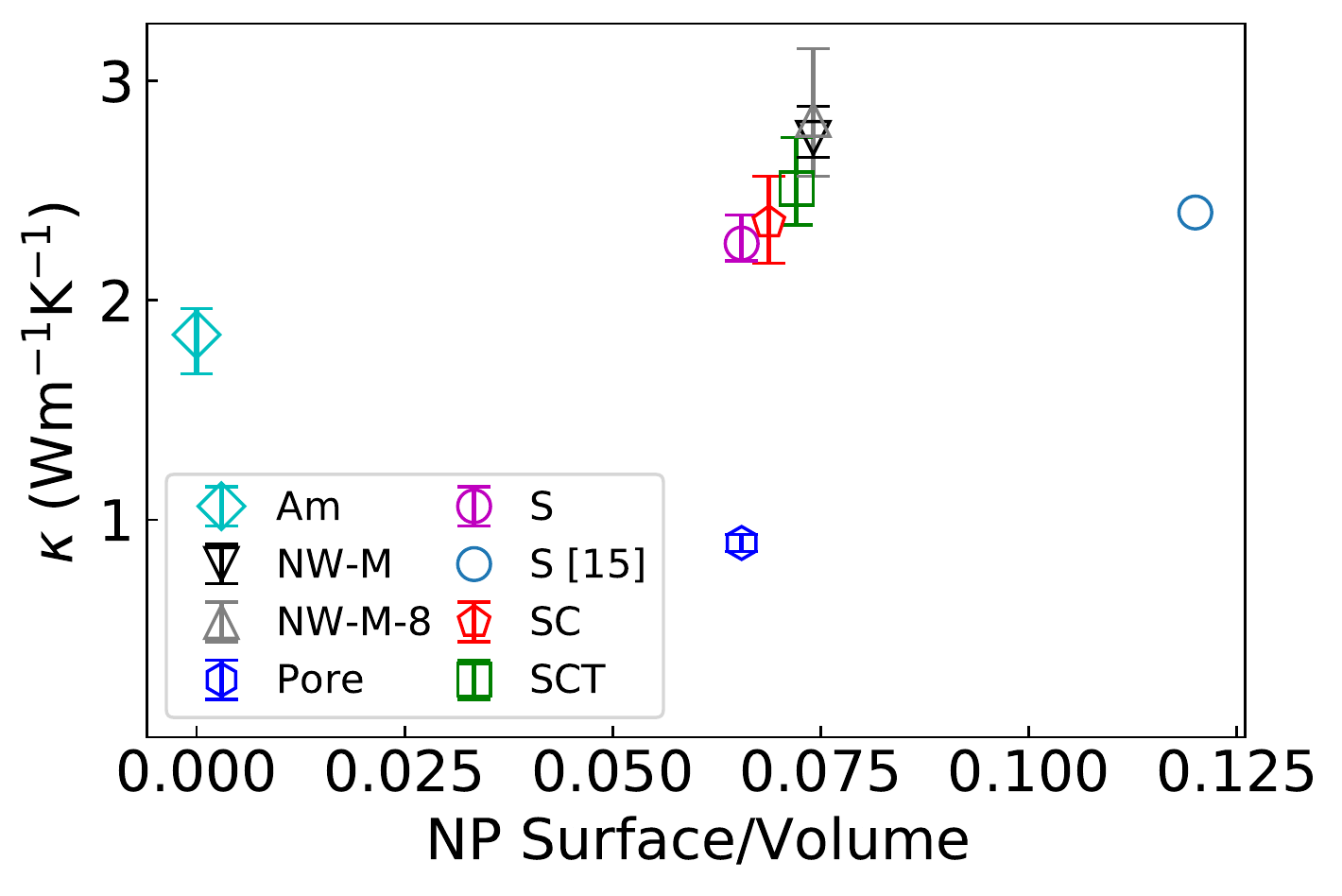}
	\caption{\label{fig:KappaVolum}Thermal conductivity estimated with EMD for the studied configurations, as a function of their surface to volume ratios. S~\cite{Tlili2019} refers to results of Tlili \textit{et al.} with the same crystalline volume fraction and NW-M-8 refers to a cube constituted by a 2x2x2 grid of the configuration NW-M.}
	
\end{figure}

The thermal conductivity can also be estimated from the Green-Kubo relation (Equation~\ref{eq:GKdis}). The results at 300~K as a function of the surface to volume ratio of the NIs are displayed in Figure~\ref{fig:KappaVolum}. A clear trend appears for the configurations containing NIs, the thermal conductivity increases with the surface to volume ratio. Moreover, the thermal conductivity is very close to one of Tlili \textit{et al}~\cite{Tlili2019} for a nanocomposite with smaller NIs representing the same volume fraction but twice the surface to volume ratio. This hints that the effect of the interconnection/structural percolation is stronger than the effect of an increased scattering surface. It is also noticeable that all the NIs of this study are regrouped in the center of the graph with ratios between \num{6.5e-2} and \num{7.5e-2}. The $\kappa$ are also very close, with intersecting error bars. In the end, only the NW-M really stands out with a $\kappa$ increased by 20\% compared to the spherical NI. Finally, the thermal conductivity for a cubic supercell of 8 NW-M (in gray) is very close to the $\kappa$ of the single NW-M. This absence of variation of the thermal conductivity shows that $\kappa$ does not depend on the number of NI simulated. 

The $\kappa$ computed for bulk a-Si through EMD is 1.9 \si{\W \per \m \per \K} which is close to previously reported values \cite{Lv2016,Larkin2014}. The nanoporous a-Si has a sub amorphous $\kappa$ due to the additional scattering at the surface of the pores. When the pores are filled with crystalline NIs, the $\kappa$ is increased by a factor 2.5 to 3 compared to the porous $\kappa$ and a factor 1.2 to 1.5 compared to a-Si.

The results of the EMD computations are compared to $\kappa_{Tot}$ obtained in the previous section in Table~\ref{tab:KComp}. As visible in Figure~\ref{fig:KappaComp300}, even if the thermal conductivity predicted by Equations~\ref{eq:kp} and \ref{eq:kd} are higher, both methods predict the same hierarchy of $\kappa$. The difference of prediction between the two methods is more pronounced for the configurations containing NIs and even more if there is a structural percolation. The last row of the table, shows the results if $\tau_{phonon-phonon}$ is not taken into account. It appears that the reduction of thermal conductivity induced by this term is marked only for the configurations with a crystalline continuity.

\begin{table}[h]
	\caption{\label{tab:KComp}Thermal conductivity in \si{\W \per \m \per \K} at 300~K for the different configurations, computed with EMD, or estimated with equations~\ref{eq:kd} and~\ref{eq:kp}. }
	\centering
	\begin{tabular}{ccccccc}
		\toprule
		&\textbf{Am}&\textbf{Pore}&\textbf{S}&\textbf{SC}&\textbf{STC}&\textbf{NW-M}\\
		\midrule
		EMD	&\num{1.9(2)}&\num{.89(4)}&\num{2.3(1)}&\num{2.3(2)}&\num{2.5(2)}&\num{2.7(1)}\\
		WP w $\tau_{ph-ph}$	&2.5&1.3&4.0&4.1&7.2&8.9\\
		WP w/o $\tau_{ph-ph}$	&2.6&1.3&4.1&4.2&8.7&14.0\\
		\bottomrule
	\end{tabular}
\end{table}

To conclude on the EMD computation, all configurations containing NIs have a $\kappa$ between 2.3 and 2.7 \si{\W \per \m \per \K} (see Table ~\ref{tab:KComp}). The NW-M is the only configuration that has a distinctively higher $\kappa$ than the nanocomposites without structural percolation. Its thermal conductivity is 20\% higher than the one of the nanocomposite with a spherical inclusion. Moreover, counter-intuitively, when going from the spherical NI to the NW-M, $\kappa$ seems to increases with the surface to volume ratio. However, usually, an increased density of interfaces leads to a reduction of $\kappa$~\cite{Minnich2007}. The increased surface to volume ratio is here a consequence of the gradual interconnection of the NIs: as the shape shifts from a sphere to a NW-M the surface to volume ratio indeed increases. In our case, the interconnection has probably a stronger effect on $\kappa$ than the surface to volume ratio. In addition, a similar thermal conductivity has been found before for smaller particles having a higher surface to volume ratio but sharing the same volume fraction than the S configuration~\cite{Tlili2019}. This means that the surface to volume ratio has little effect on c-Si NIs in a-Si matrix. This lack of impact of the surface to volume ratio contrasts with the results obtained for GaN NIs in SiO$_2$~\cite{Termentzidis2018a}. The origin of this difference may be found in the impedance mismatch between GaN and SiO$_2$.

Furthermore, the $\kappa$ of a-Si estimated here is coherent with previous results obtained with a similar method~\cite{Lv2016} and with experimental results~\cite{Cahill1994}.
Porous a-Si is less studied, however, the results can be compared with results obtained on porous c-Si amorphized by irradiation~\cite{Isaiev2014,Massoud2020}. 
The experimental results range between 3 and 1.8 \si{\W \per \m \per \K}, thus higher than the .89 \si{\W \per \m \per \K} obtained here. This difference can have multiple origins, two of which are important: the shape of the pores and the presence of gas in the pores in the experimental set up.
Given that we know the thermal conductivity of porous a-Si and of the NIs alone \cite{Fang2006}, it is tempting predicting $\kappa$ in the corresponding nanocomposite; however, a simple sum of the two thermal conductivities does not predict the correct thermal conductivity. Moreover, even an effective medium approach based on a mixing rule between the NIs and the matrix $\kappa$ is not able to predict the correct thermal conductivity.

\section{Discussion} \label{sec:Disc}

Previous analysis of the impact of NIs in amorphous matrices on the vibrational and thermal properties of nanocomposites via MD have focused on the intrinsic properties of spherical NIs and on their role as scatterers~\cite{Tlili2019,Termentzidis2018,Luo2019}. The influence of their shape and eventual interconnection are rarely the center of attention, here we try to understand their role on the effective thermal conductivity and on the ballistic transport.

\subsection{Thermal conductivity}

The gradual interconnection/structural percolation between the NIs increases the effective thermal conductivity of the studied nanocomposites. This enhancement is due to an increase of the propagative part. The diffusive part  ($\kappa_{D}$) on the contrary, is not affected by the shape of the inclusions nor by the structural percolation. It can however be noted that $\kappa_{D}$ is increased by the introduction of NIs, and that the only way to decrease it below amorphous values is to introduce pores. Finally, we showed that the two methods used to evaluate $\kappa$ conserve the same hierarchy.

Having estimated thermal conductivity through two methods (the WP method and the EMD computation), the respective results can be compared. Firstly, it appears that the two methods give slightly different results. Equations~\ref{eq:kp} and~\ref{eq:kd} of the WP method overestimate the thermal conductivity of all configurations, and particularly in case of structural percolation that makes the thermal response very inhomogeneous. It also does not take properly into account the possible thermal sensitivity of the MFP. Secondly, in opposite, the EMD simulations might not capture all the effects induced by the NIs. Additional non equilibrium molecular dynamics (NEMD)~\cite{Schelling2002b} simulations containing multiple NIs could be interesting to perform. In case of structural percolation, the heat flux will likely concentrate in the crystalline percolation and the effect of this concentration maybe lost in the flux auto-correlation over the whole sample that is used to compute $\kappa$ with the EMD method. 

However, the discrepancies between the values of both models also question the quantitative accurateness of the computation of $\kappa$ with the kinetic theory. The robustness of the method, in particular for nanocomposites, is not established. To carry out the computations, multiple assumptions are made. These different assumptions will be reviewed in section \ref{subsec:kineticDisc}.

\subsection{Ballistic and Diffusive Transport}

Concerning ballistic transport, the behavior at high and low frequencies must be distinguished. At low frequencies for the longitudinal polarization, no distinction can be made between the different nanocomposites containing NIs. The WP travels through NIs and matrix alike.
At higher frequencies, the waves are strongly attenuated in the amorphous matrix and ballistic transport is possible through the structural percolation only.

At low frequencies, ballistic propagation was expected in the amorphous matrix \cite{Beltukov2018}. Moreover, at these frequencies there is no impedance mismatch: the group velocity in a-Si and c-Si are similar (as can be seen when comparing the $v_L(\omega)$ between a-Si and c-Si in the central panel of Figure \ref{fig:KappaProp}). The long MFP at low frequencies for a-Si/c-Si nanocomposite is consistent with results obtained with finite elements simulations~\cite{Luo2019}. Moreover, the transmission rate through a a-Si/c-Si interface is known to be high for a single interface and for grain boundaries in nanocrystalline Si~\cite{Yang2017,Yang2018}.
The combination of a high MFP in the matrix, a lack of impedance mismatch and a good transmission through the interface results in a reduced impact of the NIs on the MFP at low frequencies for longitudinal polarization. 
For the transverse polarization, still at low frequencies, the MFP is similar for all the configurations, excepted for the porous one. This similarity happens in spite of the acoustic mismatch between matrix and NI (see Figure \ref{fig:KappaProp}) and the stronger scattering observed in Table \ref{tab:LFL}. The latter indicates that ballistic transport at those frequencies is dominated by the matrix and that the inclusions have little effect in spite of the distortion of the wave front.
However, it is worth mentioning that a previous study observed a decrease of MFP at low frequencies for a similar system with smaller spherical inclusions and the same crystalline volume fraction \cite{Tlili2019}. This difference might be explained by the increased density of scatterers that amplifies the interfacial effects, or by specific coherent effects as the wave-length is close to the size of the spheres in this case.
To conclude on this point, the most effective way to decrease the transmission of low frequency WP relative to bulk a-Si in these nanocomposites is to introduce pores.

While only few differences appear between the configurations at low frequencies, at high frequencies strong disparities between the nanocomposites become clear.
At high frequencies the MFP in a-Si is small \cite{Beltukov2018} and there is an impedance mismatch between a-Si and c-Si for both polarizations (see central panel of Figure \ref{fig:KappaProp}). As a result, the WP is strongly attenuated in the matrix but travels well through the structural percolation at high frequencies. A previous study has shown a similar behavior for NWs with an amorphous shell~\cite{Desmarchelier2021}. 

Interestingly, if the MFP is affected by the shape of the NIs, the diffusivity is not. All the configurations that include NIs have a very similar diffusivity. This diffusivity is distinctively higher than the bulk a-Si one (see Figure~\ref{fig:KappaDiff}). A diffusivity increase caused by the addition of NIs was already observed~\cite{Tlili2019}. The only strategy to decrease the diffusivity of a-Si seems to be the creation of pores.

To sum up on the ballistic transport properties: NIs were already known to affect the transmission of phonons, for instance, small spherical NIs act as a low pass filter~\cite{Damart2015}, and here we show that if there is a structural percolation in the nanocomposite it can be used as a bandpass filter centered at 10~THz.

\subsection{Validity of the hypothesis made}\label{subsec:kineticDisc}
Equations \ref{eq:kd} and \ref{eq:kp} rely on different hypothesis. In this section, the validity of these hypotheses along with the possible origins of the discrepancies between the models will be reviewed. 

First, both the diffusive and the propagative contributions are considered at all frequencies. In previous works the different contributions were separated either based on frequency ranges, or on the periodicity of the modes~\cite{DeAngelis2019}. Here both contributions are included for all the frequencies considered. This is motivated by the fact that both a propagative and a diffusive part appear at all the observed frequencies for our configurations (see tables~\ref{tab:HFL} and~\ref{tab:LFL}). This contributes to the overestimation of the thermal conductivity by the kinetic theory. Indeed, some modes are considered twice, once as diffusive and once as propagative. This is especially true in the low frequency range where both MFP and diffusivity are high. In such a regime the relative contributions of expressions \ref{eq:kd} and \ref{eq:kp} should be weighted.

Secondly, the propagative contribution is also very likely overestimated. This overestimation already appears for the bulk a-Si for which the propagons are expected to contribute up to 40\% of $\kappa$ \cite{Larkin2014} and our model gives 50\%. This overestimation can be attributed to the lack of a cut-off frequency for the propagative contribution as previously discussed. The effect is much more marked for the STC and NW-M nanocomposites, for those, only a small fraction of the system (restricted to the center of the crystalline part) takes part to the ballistic transport at high frequencies (see Table~\ref{tab:HFL}). The transport only happens in the structural percolation and not in the whole nanocomposite. A manifestation of this phenomenon also appears in Figure~\ref{fig:EnvEx}, part of the energy is scattered, and a part of it travels ballisticaly. The diffusive behavior is visible through the gradual flattening of the central peak (0-10~nm). The propagative behavior is given by the lobe shifting through the sample. This lobe corresponds to the WP travelling in the structural percolation. Yet, in Equation~\ref{eq:kp}, it is assumed that the whole configuration contributes to $\kappa_P$. This leads to an overestimation of $\kappa_P$, especially at high frequencies where non propagating modes are taken into account in the DOS but do not contribute to the ballistic transport.
Moreover, in the WP simulations the propagation direction is aligned with the structural percolation. This alignment decreases the interactions of the WP with the branches of the NW-M and STC perpendicular to the propagation at the crossings. This may artificially increase the MFP measured by WP propagation. This might be particularly important for the NW-M, where back scattering at intersections is expected to play an important role~\cite{Verdier2018e}, whereas no impact of the intersection of the NW is visible in Table~\ref{tab:HFL}.

For comparison purposes, the MFP can also be estimated through the DSF thanks to the damped harmonic oscillator model (see appendix~\ref{app:DHO}). 
Yet, this method is known to give lower lifetime than the estimation through WP amplitude decay rate~\cite{Beltukov2018}. Additionally, the DSF is an averaged quantity computed over the whole unit cells represented in Table~\ref{fig:Conf}. Thus, it cannot take into account the longer MFP due to transport in the structural percolation. As a result, if the MFP computed through the DSF is used to estimate $\kappa_{P}$, the hierarchy of $\kappa$ between the nanocomposites obtained with EMD is not reproduced.

Finally, the hypothesis made on the effect of temperature are important. Namely, the MFP and the diffusivity are computed at 0~K, then for $\kappa_P$ a phonon-phonon lifetime term is added to take into account the thermal effects. This phonon-phonon scattering parameter is approximated thanks to empirical coefficients derived for bulk c-Si. These coefficients were already successfully used for NWs, albeit NWs with larger characteristic dimensions than in the present study~\cite{Yang2013}. Moreover, the phonon-phonon scattering in amorphous materials is negligible. Its effect being small in front of the effect of disorder. Thus, the bulk c-Si scattering coefficients seem to be the best available. Nonetheless, these parameters might be impacted by the interfaces and size effects. Interfaces are known to increase electron-phonon coupling~\cite{Luh2002} and could as well increase the phonon-phonon scattering. The diffusivity can also be influenced by the temperature. To avoid using a temperature correction coefficients, the WP propagation simulations could be performed at higher temperature. However, at higher temperature, the amplitude of the impulsion has to be increased in order to distinguish the WP from the thermal agitation. This larger impulsion may induce other bias, as the overestimation of the effect of anharmonicity.

All those factors lead to an overestimation of the thermal conductivity computed through the kinetic theory and particularly of the propagative part.

\section{Conclusion}

The vibrational and thermal properties of gradually interconnected c-Si NIs in an a-Si matrix host matrix have been studied, with the goal of gaining a better understanding of the effects of a crystalline continuity at a constant NI volume fraction. WP simulations revealed that the structural percolation has a strong impact on the transmission of energy at high frequencies (8-12~THz), the MFP being increased by an order of magnitude in case of structural percolation. The interconnection also results in a thermal conductivity increase. This enhancement appears for the two methods used in our paper for the estimation of $\kappa$: namely WP method (kinetic theory) and EMD computation. However, the kinetic theory predicts a twofold increase of $\kappa$ between the non-interconnected NIs and the interconnected NIs, while the EMD simulations predict a more modest increase of 20\%. More generally, the use of Equations \ref{eq:kd} and \ref{eq:kp} seem to overestimate $\kappa$, especially its propagative part $\kappa_{P}$. This difference between the predictions of the two methods has multiple roots: the contribution of all frequencies to both $\kappa_{D}$ and $\kappa_{P}$, the overestimation of the MFP due to alignment effects and the incomplete consideration of temperature effects. This leads us to conclude that if ballistic transport can be observed at high frequencies for percolating NIs it does not induce a marked absolute thermal conductivity increase.
This kind of configurations could thus be used for applications where a low $\kappa$ is needed while keeping the coherent transport of phonons at high frequencies. Such properties could be useful for information processing or phonons focusing in a structure.

\appendix
\section{Damped Harmonic Oscillator}
\label{app:DHO}

The DSF can be used to extract both the dispersion relation and the lifetime of phonons. To this end the low frequency region of the DSF can be fitted with the Damped Harmonic Oscillator model~\cite{Beltukov2016}:

\begin{equation}
S_{\eta}(q,\omega)=\frac{A}{(\omega^2 - \omega^2_{\eta}(q))^2 +\omega^2\Gamma_\eta^2}
\label{eq:DHO}
\end{equation}

with $\Gamma_\eta=1/\tau_\eta$ the inverse lifetime, $\omega_{\eta}(q)$ the phonon dispersion, and $A$ the amplitude and $\eta$ labeling either the longitudinal or transverse polarization. The parameters of Equation~\ref{eq:DHO} are fitted to match the DSF obtained with Equation \ref{eq:DSF} for every wave vector. This fit is realized on the DSF convoluted with the experimental resolution~\cite{Damart2015}. Thus, this model enables the computation of both the lifetime and the dispersion relation using the DSF only. 
\begin{figure}[h]
	\centering
	\includegraphics[width=.5\textwidth]{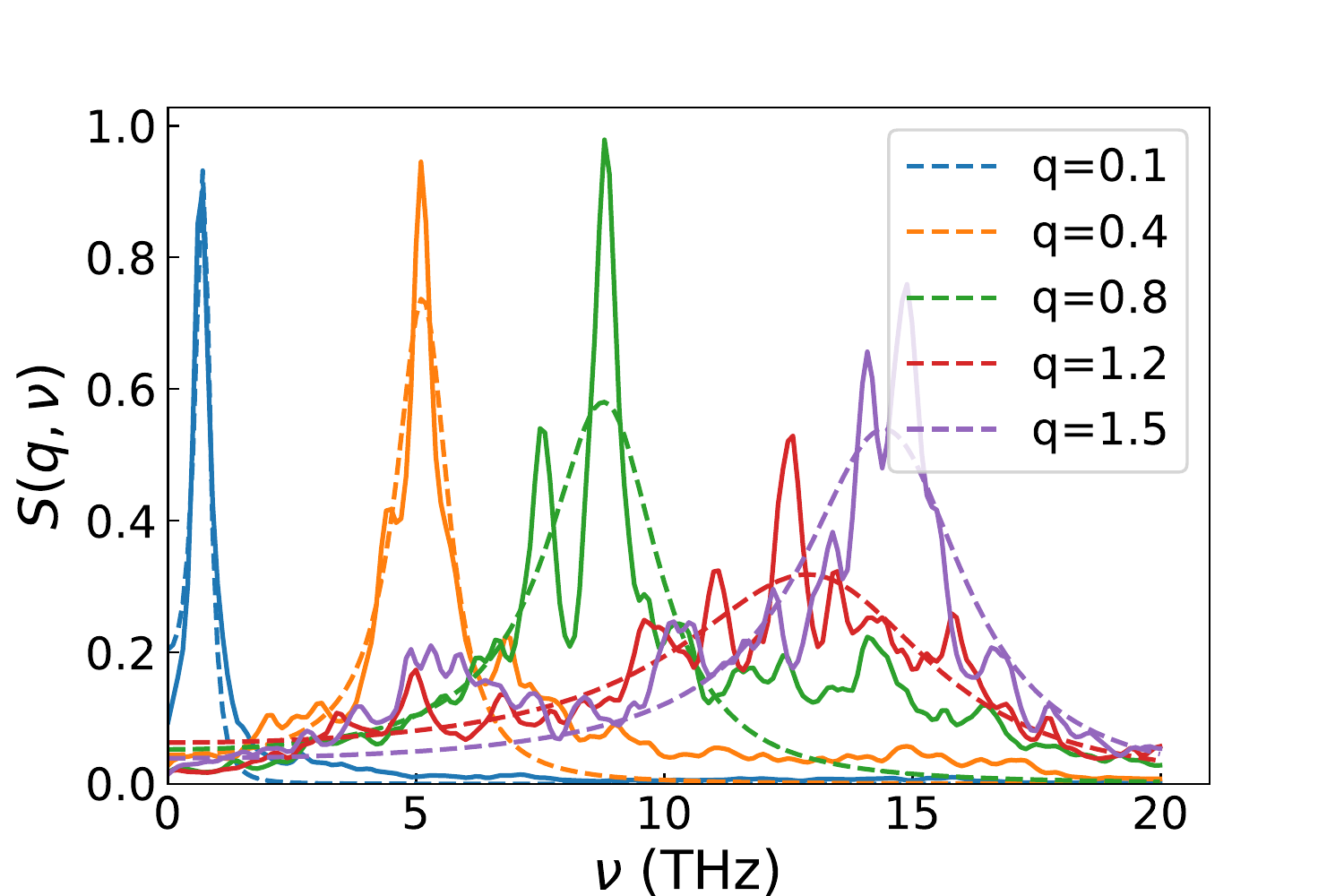}\includegraphics[width=.5\textwidth]{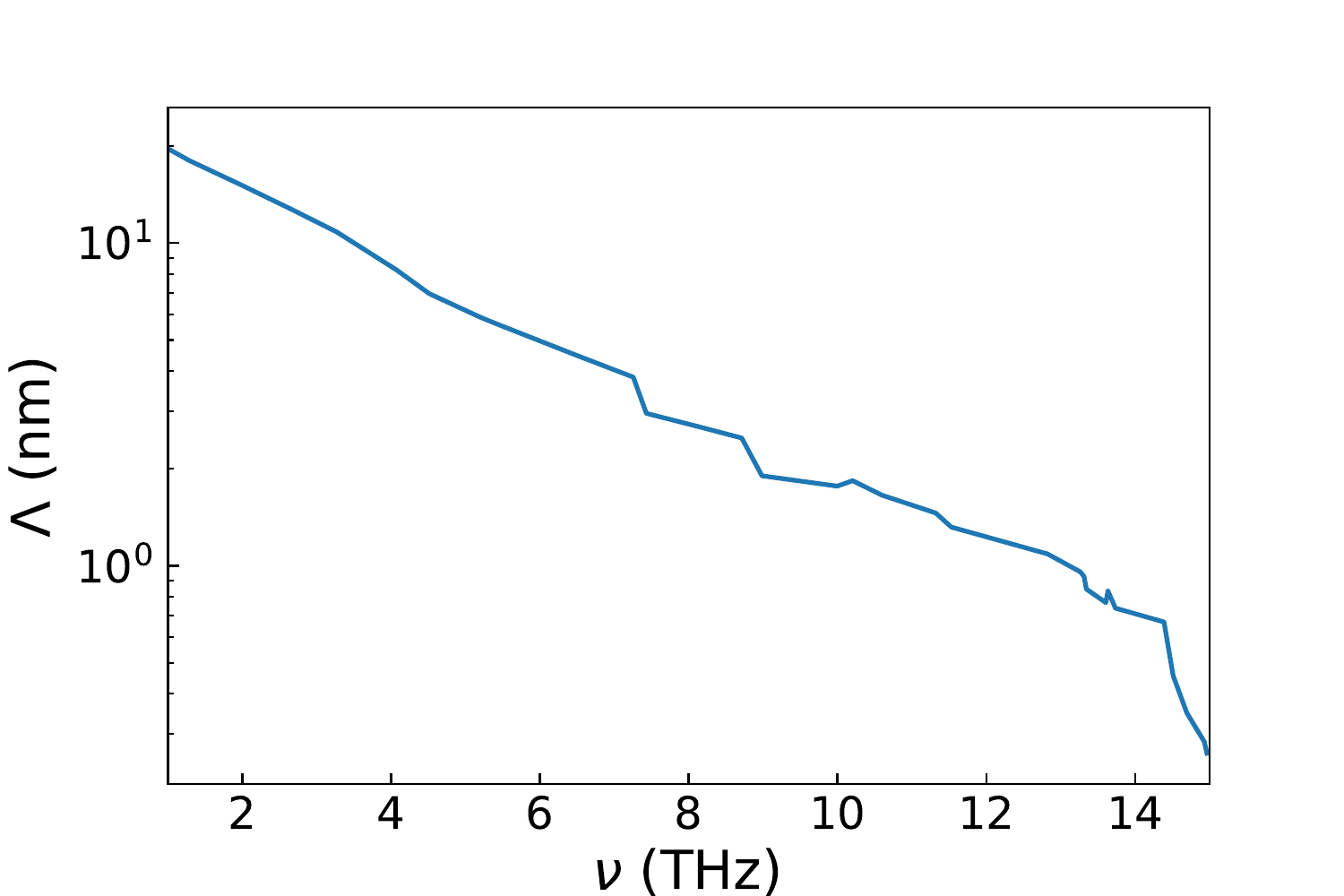} 
	\caption{\label{fig:DHO} $S_{\eta}(q,\omega)$ as fitted from Equation~\ref{eq:DHO} (left panel) and resulting MFP (right panel) for the NW-M configuration for the longitudinal polarization. }
\end{figure} 
The result of the fitting to a DHO model for a longitudinal polarization for the NW-M is displayed in the left panel of Figure~\ref{fig:DHO} for a few wave vectors. The expression~\ref{eq:DHO} seems to fit to the DSF computed via Equation~\ref{eq:DSF} reasonably well at low frequencies. However, as the frequency increases the DSF is increasingly noisy, degrading the fit quality. The MFP can also be extracted (left panel of figure~\ref{fig:DHO}). It then appears that the MFP peak at 10~THz observed with the WP disappears. The MFP computed with this method steadily decreases. This discrepancy can be attributed to the low frequency limit of validity of this fit discussed in~\cite{Beltukov2016} or to the fact that this MFP is computed using a quantity averaged over the whole sample. This spatial average cannot capture the effect of the structural percolation showcased with the WP method. 

The successive fits of different wave-vector enable the computation of the dispersion relations $\omega_{\eta}(q)$. Dispersion relations from which the group velocity can be computed. However, the method includes more parameter to fit than the method described in section~\ref{sec:DSF} and is thus a less robust approach.

\section{DOS estimated with DSF, VACF or KPM } \label{app:VDOSDSF}

The VDOS can be estimated thanks to the integral of the DSF over the wave vectors. This allows a distinction between the different wave-vector directions and polarizations. To consider the anisotropy of c-Si the sum for the different vector direction of the Brillouin zone (BZ) are considered. The VDOS obtained are then filtered using Savitzky-Golay polynomial filter. 

\begin{figure}[h]
	\centering
	\includegraphics[width=\textwidth]{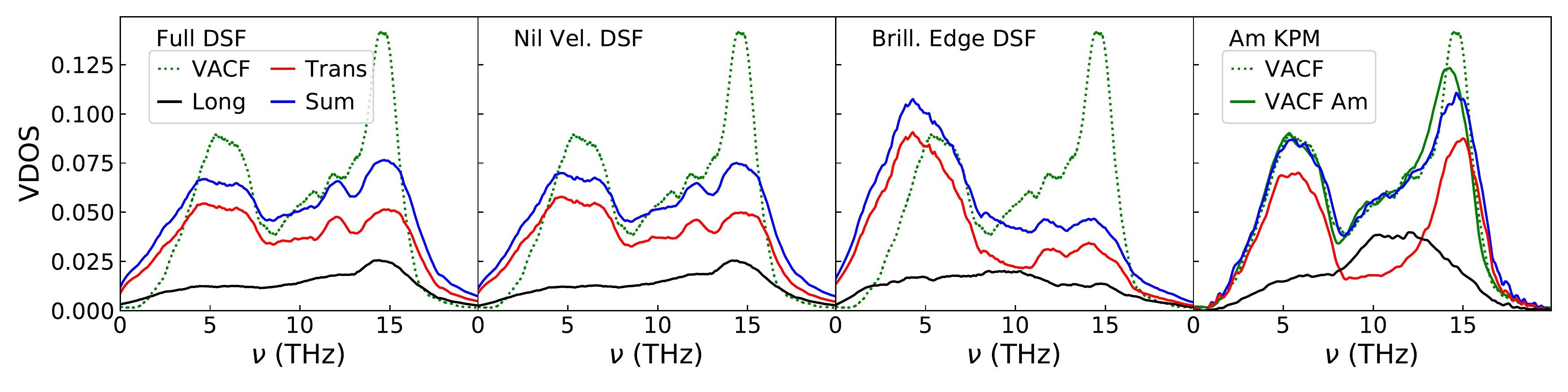}
	\caption{\label{fig:DOSDSF} VDOS computed using different methods, the red and blue lines correspond respectively to the transverse and longitudinal polarizations, the solid blue and green dashed lines correspond to the total VDOS computed respectively via the DSF or the VACF. The first panel starting from the left uses the full DSF, the second the DSF limited to the theoretical wave-vector limit of the BZ of c-Si, the third the DSF limited to the wave vector giving a 0 velocity and the last displays the results using the KPM method for an amorphous sample.}
\end{figure}

However, a choice has to be made regarding the range of wave-vector considered for the integration. The different possibilities tested are displayed in figure~\ref{fig:DOSDSF}. They are compared with the results of the VACF, which will serve as a reference point. 
The first possibility is to consider the whole DSF computed, that is from 0 up to 2.25 \si{\per \angstrom} ("Full DSF" in figure~\ref{fig:DOSDSF}). Another possibility is to consider the point at which the estimated group velocity of the phonons is nil ("Nil Vel. DSF" in the figure~), relying on the fact that near the BZ limit the velocity is nil. The last possibility is to consider the theoretical end of the BZ computed for the lattice parameter used ("Brill. Edge DSF" in figure~\ref{fig:DOSDSF}). Another comparison point for the study is the separation of transverse and longitudinal VDOS by the KPM method~\cite{Beltukov2015} on a fully amorphous sample, given that the configurations are mostly amorphous (70\% of a-Si and 30\% of c-Si).

When comparing the different VDOS it appears that the total VDOS computed using DSF does not match with the VACF results, whatever the wave vector limit used. If all the peaks seem to be present, their relative size does not match, and they are flattened. This is even more marked when considering the theoretical end of the BZ (third panel in Figure~\ref{fig:DOSDSF}). The failure to reproduce the VDOS using the DSF can partially be attributed to the fact that the BZ is not defined for the amorphous Si. 

In the other hand, the VDOS of a-Si of computed via the KPM method matches comparatively well the VDOS of the NW-M configuration computed via VACF. Still, in this case some differences arise after 12~THz, but given the reduced MFP at those frequencies, its impact on the computation of $\kappa_{P}$ is negligible.

\bibliographystyle{ieeetr}
\bibliography{references}  






\end{document}